\documentclass[]{elsarticle}

\usepackage{multicol}

\usepackage{amsmath}
\usepackage{stmaryrd}

\usepackage[skip=5pt plus1pt, indent=40pt]{parskip}

\usepackage{hyperref} 
\usepackage{calc}
\usepackage{graphicx}
\usepackage{graphics}
\graphicspath{{./Figures/}}
\usepackage{pdflscape}
\usepackage{adjustbox}
\usepackage[ampersand]{easylist}
\ListProperties(Indent2=0.5cm,Indent3=1cm,Indent4=1.5cm,Indent5=2cm)
\usepackage{caption}
\usepackage{comment}
\usepackage{multirow}
\usepackage{units}
\usepackage{booktabs}
\usepackage[T1]{fontenc} 
\usepackage{csquotes}
\usepackage{import}
\usepackage{mathrsfs}
\usepackage{amssymb}
\usepackage[overload]{empheq}
\usepackage{aliascnt}
\usepackage{cleveref}
	\crefname{equation}{}{}
\usepackage{dirtree}
\usepackage{relsize} 
\usepackage{enumitem} 
\newlist{steps}{enumerate}{1} 
\setlist[steps, 1]{label = Step \arabic*:} 

\usepackage{menukeys}

\usepackage{multirow} 
\usepackage{array} 
\usepackage{float} 

\usepackage{anyfontsize}
\usepackage{listings}
\usepackage{tikz}
\usepackage[caption=false]{subfig}

\usepackage{geometry}
\usepackage{todonotes}

\usepackage{cases}
\usepackage[edges]{forest}
\usepackage{fontawesome} 
\definecolor{folderbg}{RGB}{124,166,198}
\definecolor{folderborder}{RGB}{110,144,169}
\newlength\Size
\tikzset{%
  folder/.pic={ 
      \draw [pic actions] (1ex,-1ex) -- ++(0,2ex) -- ++(.25ex,.25ex) -- ++(1.25ex,0) -- ++(.25ex,-.3ex) --  ++(-1.8ex,0) -- ++(3ex,0)  |- cycle;}}

\usepackage{xurl}
\biboptions{sort&compress}

\usepackage{macros/finiteVolume}
\usepackage{macros/tensorNotation}
\input{./macros/subequations.tex}

\usepackage{manyfoot}%

\begin{document}

\begin{frontmatter}

\title{multiRegionFoam -- A Unified Multiphysics Framework for Multi-Region Coupled Continuum-Physical Problems}

\author[address1]{Heba Alkafri}
\author[address1]{Constantin Habes \corref{correspondingauthor1}}
\author[address1]{Mohammed Elwardi Fadeli}
\author[address2]{Steffen Hess}
\author[address2,address3]{Steven B. Beale}
\author[address2]{Shidong Zhang}
\author[address4]{Hrvoje Jasak}
\author[address1]{Holger Marschall \corref{correspondingauthor2}}

\cortext[correspondingauthor1]{Corresponding author, Email: constantin.habes@tu-darmstadt.de}
\cortext[correspondingauthor2]{Corresponding author, Email: holger.marschall@tu-darmstadt.de}

\address[address1]{Technische Universität Darmstadt, Department of Mathematics, Computational Multiphase Flow, 64287 Darmstadt, Germany}
\address[address2]{Forschungszentrum Jülich GmbH, Institute of Energy and Climate Research (IEK-9, IEK-13, IEK-14), 52428 Jülich, Germany}
\address[address3]{Queen's University, Department of Mechanical and Materials Engineering, Kingston K7L 3N6, Ontario, Canada}
\address[address4]{The Cavendish Laboratory, Department of Physics, University of Cambridge, United Kingdom}

\begin{abstract}
This paper presents a unified framework, called multiRegionFoam, for solving multiphysics problems of the multi-region coupling type within OpenFOAM (FOAM-extend). It is intended to supersede the existing solver with the same name. The design of the new framework is modular, allowing users to assemble a multiphysics problem region-by-region and coupling conditions interface-by-interface. The present approach allows users to choose between deploying either monolithic or partitioned interface coupling for each individual transport equation. The formulation of boundary conditions is generalised in the sense that their implementation is based on the mathematical jump/transmission conditions in the most general form for tensors of any rank. 

The present contribution focuses on the underlying mathematical model for these types of multiphysics problems, as well as on the software design and resulting code structure that enable a flexible and modular approach. Finally, deployment for different multi-region coupling cases is demonstrated, including conjugate heat, multiphase flows and fuel-cells.

Source code repository:  \url{https://bitbucket.org/hmarschall/multiregionfoam/}

\end{abstract}




\end{frontmatter}




\section{Introduction}

Interface-coupled multi-region problems like fluid-structure interactions, conjugate heat \& mass transfer or multiphase flow problems represent a subgroup of multiphysics problems of high relevance in engineering. Analysis of the underlying continuum-physical models reveals inherent structural similarities, which can be exploited in software design and method development to devise a unified computational multiphysics framework for multi-region coupled continuum-physical problems over a broad spectrum of applications.

\noindent
We have developed a novel unified solver framework for computational multiphysics of multi-region coupling type, i.e.\ transport processes coupled across region boundaries/interfaces. The code design is such that a multiphysics problem can be assembled region-by-region, and the coupling conditions interface-by-interface. Both monolithic and partitioned coupling for each individual transport equation can be applied as desired by the user. Fluid flow problems are dealt with using SIMPLE, PISO and PIMPLE pressure-velocity algorithms -- with loops of predictor and corrector steps across regions. The code is implemented as a C++ library in OpenFOAM (FOAM-extend) for computational continuum physics \cite{weller_tensorial_1998}  and follows the principles of object-oriented programming.

\noindent
We incorporate user-defined types of regions representing sub-domains of specific physics i.e.\ a set of transport equations. Such a region type should govern a meaningful subset of physics specific to the region -- such as e.g.\ fluid flow, solid mechanics, species and/or energy transport -- and can be combined with others. This results into a modular concept and allows to assemble a multiphysics problem region-by-region. The coupling and communication between regions is realised in a modular fashion where interface-specific physics as well as interpolation/mapping methods are accessible in boundary conditions. The implementation is readily parallelised for large scale computations in domain decomposition mode for runs on distributed-memory parallel computer architectures. 

\subsection*{Literature Survey}

There are numerous open-source solutions to cope with multi-region coupled problems. However, the majority of codes are dedicated to specific problems, and thus are following a domain-driven design. In consequence, they cannot be easily adapted to other multi-region coupling problems. 
Available proprietary simulation codes, on the other-hand-side, often do provide platforms to solve for a broad range of multi-region coupling problems. However, being proprietary the source-codes are non-accessible to the community with the consequence of limited flexibility and/or extensibility, particularly when it comes to very specific engineering applications in technology niches. 
To alleviate these limitations, (mostly) open-source multi-code coupling approaches have been devised. Examples are the \texttt{ADVENTURE\_Coupler} (ADVanced ENgineering analysis Tool for Ultra large REal world) \cite{kataoka_parallel_2014}, \texttt{MpCCI} (Mesh-based parallel Code Coupling Interface) \cite{noauthor_mpcci_nodate, joppich_mpccitool_2005}, \texttt{OpenPALM} (Projet d'Assimilation par Logiciel Multimethodes) \cite{buis_palm_2005, duchaine_analysis_2015}, the \texttt{OASIS} coupler \cite{valcke_oasis3_2013, craig_development_2017}, \texttt{PIKE} (Physics Integration KErnerls) \cite{pawlowski_physics_2014} as a part of the Trilinos library \cite{the_trilinos_project_team_trilinos_nodate}, and \texttt{preCICE} \cite{bungartz_precice_2016, noauthor_precice_nodate}.
With these coupling software packages, multiple distinct simulations codes are coupled in a co-simulation run -- each with an own specialisation coping with the physics in one specific region of the multi-region domain. Note that such code-to-code coupling frameworks for co-simulations are particularly not within the scope of this work, since they inherently only provide partitioned coupling strategies and thus suffer from stability and/or efficiency issues when it comes to challenging, e.g.\ numerically stiff, coupling problems.

\noindent
In what follows, we attempt to provide a comprehensive yet concise overview over available open-source multi-region coupling software in the literature, highlighting their coverage of applications.

\begin{description}

\item[Alya] \cite{casoni_alya_2015, noauthor_alya_nodate} 
is a Fortran and C based code developed at Barcelona Supercomputing Center, Spain. It solves coupled multiphysics problems using high performance computing techniques for distributed and shared memory supercomputers. The simulations involve the solution of partial differential equations in an unstructured mesh using finite element methods. Indeed it provides region coupling in a single code environment as well as partitioned multi-code coupling using existing code. Examples of implementation for fluid-structure interaction (FSI) and cojugate heat transfer (CHT) problems, among other applications, are found in  \cite{cajas_fluid-structure_2018} and \cite{casoni_alya_2015}.

\item[code\_saturne] \cite{noauthor_code_saturne_nodate} 
is developed primarily by Électricité de France R\&D (EDF) for computational fluid dynamics (CFD) applications. It is written in C and Fortran and relies on finite volume discretisation. It can be coupled with other codes, using its Parallel and Locator Exchange library (PLE) \cite{noauthor_ple_nodate}, for instance, with SYRTHES \cite{noauthor_syrthes_nodate}, a code for transient thermal simulations in solids, to model CHT problems  \cite{houzeaux_partitioned_2017}. It also has a module for arbitrary Lagrangian Eulerian (ALE) interface tracking in the frame of fluid-structure interaction \cite{houbar_simulation_2021}.

\item[deal.II] \cite{noauthor_deal.II_nodate} is a C++ library intended to serve as an interface to the complex data structures and algorithms required for solving partial differential equations using adaptive finite element methods. It is deployed in multiphysics simulations for various applications including FSI in ALE formulation \cite{wick_2013_solving}, and numerous others \cite{noauthor_deal.II_publications_nodate}.

\item[FEniCS] \cite{noauthor_fenics_nodate} is a python and C++ based code dedicated to solving partial differential equations arising in scientific models using the finite element method. It is often integrated with other platforms as it does not have a built in multi-physics solver. Example of usage include the FEniCS-FEATool solver \cite{noauthor_fenicsFEAtool_nodate}, the coupling of FEniCS with OpenFOAM through the multiscale universal interface MUI \cite{liu_high-performance_2022}, as well as the combination of FEnics with HAZmath \cite{Budisa_HAZnics} where the \texttt{cbc.block} extension is used that enables the assembly and the solution of block-partitioned problems.

\item[MOOSE] stands for Multiphysics Object-Oriented Simulation Environment \cite{noauthor_moose_nodate}. It is an open-source C++ based code developed at the Idaho National Laboratory, enabling parallel multiphysics simulation. It uses a finite element framework and supports segregated and fully implicit volumetric coupling as well as partitioned interface coupling. Fluid dynamics, heat transfer, and fluid–structure interaction are some of the applications where MOOSE is used \cite{permann_moose_2020, dhulipala_development_2022}.
\item[OpenFOAM] \cite{weller_tensorial_1998} (Open Field Operation And Manipulation) is an open-source C++ library for computational continuum physics (CCP) including computational fluid dynamics (CFD) based on the finite volume method (FVM) with support for dynamic meshes of general topology (unstructured meshes).
Within OpenFOAM, numerous application-specific multi-region frameworks are available mostly from its active developer community. For instance, \texttt{solids4foam} \cite{cardiff_open-source_2018} has been developed for FSI simulations, \texttt{openFuelCell} \cite{beale_open-source_2016} for modelling fuel cells, \texttt{chtMultiRegionFoam} \cite{renze_simulation_2019} for CHT problems.
Note that besides the above application-specific solutions to interface-coupled multiphysics, OpenFOAM (OpenFOAM-dev) has undergone refactoring towards a "new modular solver framework" \cite{noauthor_NewModular_nodate}, in which so-called solver modules can be selected for coupled multi-region simulations. However, with only one solver module selectable for each region, the resulting framework forces the user to develop modules covering the full set of physics in a region. This hinders flexibility and introduces unnecessary complexity. Moreover, interfacial physics has not been modularised and generalised at all. Both aspects have been subject to the present work.

\item[Yales2] \cite{noauthor_yales2_nodate} 
aims at solving two-phase combustion from primary atomization to pollutant prediction on massive complex meshes. It is developed at CORIA-CFD using C++ and Fortran. It uses a finite volume solver for multiphysics problems in fluid dynamics, with support for ALE within FSI context \cite{sarkar_mechanism_2021} in addition to the possibility for multi-code coupling, such as for CHT applications \cite{boulet_modeling_2018} through the OpenPALM library \cite{noauthor_openpalm_nodate}.
\end{description}
\noindent
While the focus here is on open-source, there are also various proprietary software packages developed for similar purposes, such as COMSOL Multiphysics \cite{noauthor_comsol_nodate}, FEATool \cite{noauthor_featool_nodate}, Fluent \cite{noauthor_ansys_nodate-1}, and LS-DYNA \cite{noauthor_ansys_nodate}.
\subsection*{Aim \& Objective}
This contribution aims to provide a unified and versatile framework to cope with multiphysics problems of multi-region coupling type in OpenFOAM. Particular emphasis is put on
\begin{itemize}
    \item freedom in the choice of coupling strategies -- i.e.\ monolithic and partitioned coupling is at the choice of the user for each individual transport equation,
    \item ease and flexibility in assembling multiphysics problems -- by means of use-defined region types, which also can be superimposed, leading to a multiphysics setup that can be assembled in a modular fashion,
    \item support of established predictor-corrector based solution algorithms -- e.g.\ to support pressure-velocity, or magnetohydrodynamics in fluid flow across regions,
    \item strict physics-driven design which allows to clearly separate material models from balance equations -- by rigorously exploiting the common mathematical structure of transport equations \emph{and} interface jump and transmission (flux) conditions.
\end{itemize}
With this, it is hoped to enable substantial coverage over a spectrum of different multiphysics problems of multi-region coupling type, and to leverage significant synergies among different, so far disjoint domain-expert communities, such that improvements and fixes from one community directly can benefit others.

\noindent
In the remainder, detailed information on the generic mathematical formulation of the sharp-interface model is given (Section \ref{sec:mathModel}) which is exploited at the software design stage in the code structure to arrive at a unified multiphysics framework (Section \ref{sec:mutliRegionFoam}). Eventually, its deployment for different multi-region coupling cases is demonstrated (Section \ref{sec:usage}).

\section{Generic sharp interface model} \label{sec:mathModel}

We aim to provide a concise self-contained derivation of the sharp interface model in its generic formulation as it emerges from balance considerations of conserved quantities. The mathematical procedure will also yield the well-known general transport equation introduced by Spalding in \cite{a_d_gosman_heat_1969}. Thus, the following can also be seen as its extension to general transport equations in multiple domains coupled across sharp interfaces separating the domains. For this, we shall closely follow \cite{bothe_sharp-interface_2022, slattery_interfacial_2007, ishii_thermo-fluid_2011, kjelstrup_non-equilibrium_2008}.

\begin{figure}
  \centering
  \def\svgscale{0.7}
  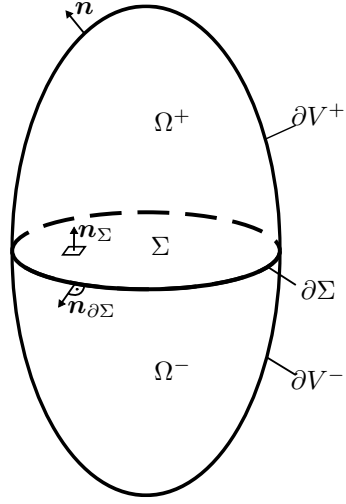
  \caption{Material volume $V$ with two subdomains $\Omega^{+}$ and $\Omega^{-}$ separated by a sharp interface $\Sigma$}
  \label{fig:Sharp_interface_egg}
\end{figure}

\noindent
Starting point of this derivation is a conservation equation in its generic form,
\begin{equation} \label{eqn:General_Conservation_Equation_1}
      \frac{D \Phi}{D t}
    =
      J + S,
\end{equation}
with the left hand side of \eqref{eqn:General_Conservation_Equation_1} being the material derivative of an extensive quantity $\Phi$ (e.g.\ mass, momentum, energy), $J$ denoting the flux term and $S$ being sources/sinks of $\Phi$.
We assume this generic conservation equation to be valid in a material control volume $V(t)=\Omega^{+}(t) \cup \Omega^{-}(t)$, being composed out of two subdomains $\Omega^{+}$ and $\Omega^{-}$ (see Fig.\ \ref{fig:Sharp_interface_egg}). The two subdomains are separated by a deformable sharp interface $\Sigma(t)=
\partial \Omega^{+}(t) \cap \partial \Omega^{-}(t)$
which leads to the fact that V(t) is bounded by
$\partial V(t) = \partial V^{+}(t) \cup \partial V^{-}(t) \cup \partial \Sigma(t)$, where $\partial V^{\pm}(t) = \partial \Omega^{\pm}(t) \setminus \Sigma(t)$.
Furthermore, we define that the interface normal $\boldsymbol{n}_\Sigma$ always points from $\Omega^{-}$ to $\Omega^{+}$ and the interface edge normal points out of $\Sigma$.

\noindent
The sharp interface $\Sigma(t)$ considered here can be seen as a simplification of a transition layer of finite thickness between adjacent domains with different physical properties. Therefore, the interface is not necessarily mass-less and can store conserved quantities which are accounted for in the following by the appearance of so-called surface excess quantities \cite{bothe_sharp-interface_2022}.
Thus, the extensive quantity can be written as a volume integral, 
\begin{equation} \label{eqn:General_extensive_quantity}
      \Phi 
    = 
      \int_{V(t)} \rho\phi\mathrm{d} \boldsymbol{x}
    + \int_{\Sigma(t)} \rho^{\Sigma} \phi^{\Sigma} \mathrm{d} \boldsymbol{x}
    \; ,
\end{equation}
where $\rho$ is the mass density in $\Omega^\pm$, $\phi$ is the mass density-related volume specific density of the extensive quantity in the bulk and $\rho^{\Sigma}$ and $\phi^{\Sigma}$ are their respective area specific excess quantities defined on the interface \cite{slattery_interfacial_2007}. 
The flux term can be expressed through 
\begin{equation} \label{eqn:General_surface_flux}
      J 
    =
    - \int_{\partial V(t)} \boldsymbol{j} \cdot \boldsymbol{n} \mathrm{d} s
    - \int_{\partial \Sigma(t)} \boldsymbol{j}^{\Sigma} \cdot \boldsymbol{n}_{\partial\Sigma} \mathrm{d} l
    \; ,
\end{equation}
with the first integral being a surface integral over the bulk flux density $\boldsymbol{j}$ across the boundary of $V(t)$ and the second integral being a line integral over the interface flux density $\boldsymbol{j}^{\Sigma}$ across the interface boundary. 
A similar expression as (\ref{eqn:General_extensive_quantity}) can be formulated for the source/sink term
\begin{equation} \label{eqn:General_source_sink}
      S
    =
      \int_{V(t)} s \mathrm{d} \boldsymbol{x}
    + \int_{\Sigma(t)} s^{\Sigma} \mathrm{d} s
    \;.
\end{equation}
Here $s$ is the source/sink density field in the bulk and $s^{\Sigma}$ is its respective counterpart on the interface. Inserting (\ref{eqn:General_extensive_quantity}), (\ref{eqn:General_surface_flux}) and (\ref{eqn:General_source_sink}) into (\ref{eqn:General_Conservation_Equation_1}) gives
\begin{equation}
\label{eqn:General_Conservation_Equation_2}
    \frac{D}{Dt}\int_{V(t)} \rho\phi \mathrm{d} \boldsymbol{x} 
    +
    \frac{D}{Dt}\int_{\Sigma(t)} \rho^{\Sigma}\phi^{\Sigma} \mathrm{d} s
    = 
    - \int_{\partial V(t)} \boldsymbol{j} \cdot \boldsymbol{n} \mathrm{d} s
    - \int_{\partial \Sigma(t)} \boldsymbol{j}^{\Sigma} \cdot \boldsymbol{n}_{\partial\Sigma} \mathrm{d} l 
    \quad \, + \int_{V(t)\setminus\Sigma(t)} s \mathrm{d} \boldsymbol{x}
    + \int_{\Sigma(t)} s^{\Sigma} \mathrm{d} s\,. 
\end{equation}
The generalized transport theorem, \cite{casey_derivation_2011}
\begin{equation}
\label{eqn:Two_phase_rtt}
\nonumber
      \frac{D}{Dt}\int_{V(t)} \rho\phi \mathrm{d} \boldsymbol{x} 
    =
      \int_{V(t)\setminus\Sigma(t)}\frac{\partial \rho\phi}{\partial t} \mathrm{d} \boldsymbol{x}
    + \int_{V(t)\setminus\Sigma(t)}\nabla\cdot(\rho \phi \boldsymbol{u}) \mathrm{d} \boldsymbol{x} 
    + \int_{\Sigma(t)} \llbracket \rho \phi (\boldsymbol{u} - \boldsymbol{u} ^{\Sigma}) \rrbracket \cdot \boldsymbol{n}_{\Sigma} \mathrm{d} s\,,
\end{equation}
can be applied to the first term on the left hand side of (\ref{eqn:General_Conservation_Equation_2}), where the jump brackets are defined as 
\begin{equation}\label{eqn:Jump_Bracket_Definition}
      \llbracket \phi \rrbracket(t, \boldsymbol{x}) 
    =
      \lim _{h \rightarrow 0^{+}} 
      \left[ \phi\left(t, \boldsymbol{x}+h \boldsymbol{n}_{\Sigma}\right)
    - \phi\left(t, \boldsymbol{x}-h \boldsymbol{n}_{\Sigma}\right) \right], \quad \boldsymbol{x} \in \Sigma
    \;.
\end{equation}
Furthermore, $\boldsymbol{u}$ is the velocity field in the bulk and $\boldsymbol{u}^{\Sigma}$ is the interface velocity field which -- in the general case of a fluid interface -- can have contributions in both normal and tangential direction to the interface. 
The second term on the left hand side of (\ref{eqn:General_Conservation_Equation_2}) can be reformulated through the use of the surface transport theorem \cite{slattery_interfacial_2007}
\begin{equation} \label{eqn:Interface_tt}
      \frac{D}{Dt}\int_{\Sigma(t)} \rho^{\Sigma}\phi^{\Sigma} \mathrm{d} s
    =
      \int_{\Sigma(t)}\frac{\partial \rho^{\Sigma}\phi^{\Sigma}}{\partial t} \mathrm{d} s
    + \int_{\Sigma(t)}\nabla_{\Sigma}\cdot(\rho^{\Sigma} \phi^{\Sigma} \boldsymbol{u}^{\Sigma}) \mathrm{d} s
    \; .
\end{equation}
Here the interface Nabla operator is defined by $\nabla_{\Sigma} = \left[\mathbf{I}-\boldsymbol{n}_{\Sigma} \otimes \boldsymbol{n}_{\Sigma}\right] \cdot \nabla$,
with interface divergence of a vector $\nabla_{\Sigma}\cdot \boldsymbol{y} = tr(\nabla_{\Sigma} \boldsymbol{y})$
being the trace of the interface gradient \cite{ishii_thermo-fluid_2011}.
Using the two-phase divergence theorem \cite{slattery_interfacial_2007}, the first term on the right hand side of (\ref{eqn:General_Conservation_Equation_2}) can be written as
\begin{equation} \label{eqn:Two-phase_divergence_t}
    - \int_{\partial V(t)} \boldsymbol{j} \cdot \boldsymbol{n} \mathrm{d} s
    =
    - \int_{V(t)\setminus\Sigma(t)} \nabla \cdot \boldsymbol{j} \mathrm{d} \boldsymbol{x}
    - \int_{\Sigma(t)} \llbracket \boldsymbol{j} \rrbracket \cdot \boldsymbol{n}_{\Sigma} \mathrm{d} s
    \; .
\end{equation}
Similarly, the second term on the right hand side of (\ref{eqn:General_Conservation_Equation_2}) can be expressed using the surface divergence theorem \cite{cermelli_transport_2005}
\begin{equation} \label{eqn:Surface_divergence_t}
    - \int_{\partial \Sigma(t)} \boldsymbol{j}^{\Sigma} \cdot \boldsymbol{n}_{\partial\Sigma} \mathrm{d} l
    =
    - \int_{\Sigma(t)} \nabla_{\Sigma} \cdot \boldsymbol{j}^{\Sigma} \mathrm{d} s
    \; .
\end{equation}
Substituting (\ref{eqn:Two_phase_rtt}), (\ref{eqn:Interface_tt}), (\ref{eqn:Two-phase_divergence_t}) and (\ref{eqn:Surface_divergence_t}) back into (\ref{eqn:General_Conservation_Equation_2}) yields
\begin{equation} \label{eqn:General_Conservation_Equation_3}
    \begin{aligned} 
          &   \int_{V(t)\setminus\Sigma(t)} \left[
              \frac{\partial (\rho\phi)}{\partial t}
            + \nabla\cdot(\rho \phi \boldsymbol{u})
            + \nabla \cdot \boldsymbol{j} - s 
          \right] \mathrm{d} \boldsymbol{x} \\
        + &   \int_{\Sigma(t)} \left[
              \frac{\partial (\rho^{\Sigma}\phi^{\Sigma)}}{\partial t} 
            + \nabla_{\Sigma}\cdot(\rho^{\Sigma} \phi^{\Sigma} \boldsymbol{u}^{\Sigma})
            + \nabla_{\Sigma} \cdot \boldsymbol{j}^{\Sigma}
            + \llbracket \rho \phi (\boldsymbol{u} - \boldsymbol{u} ^{\Sigma}) +\boldsymbol{j} \rrbracket \cdot \boldsymbol{n}_{\Sigma}
            - s^{\Sigma}
            \right] \mathrm{d} s
        = 0
        \; .
    \end{aligned}
\end{equation}
Localizing this expression to points inside the bulk results in the well-known general transport equation 
\begin{equation} \label{eqn:Local_Bulk_Conservation}
      \frac{\partial \rho\phi}{\partial t}
    + \nabla\cdot(\rho \phi \boldsymbol{u})
    = 
    - \nabla \cdot \boldsymbol{j} 
    + s
\end{equation}
introduced by Spalding \cite{a_d_gosman_heat_1969}. 
Doing the same for points on the interface yields 
\begin{equation} \label{eqn:Local_Interface_Transport}
      \frac{\partial \left(\rho^{\Sigma}\phi^{\Sigma}\right)}{\partial t} 
    + \nabla_{\Sigma}\cdot(\rho^{\Sigma} \phi^{\Sigma} \boldsymbol{u}^{\Sigma})
    + \llbracket \rho \phi (\boldsymbol{u} - \boldsymbol{u} ^{\Sigma}) +\boldsymbol{j} \rrbracket \cdot \boldsymbol{n}_{\Sigma}
    = 
    - \nabla_{\Sigma} \cdot \boldsymbol{j}^{\Sigma}
    + s^{\Sigma}
    \;,
\end{equation}
which is called the interface transmission or flux condition. This equation relates the transport of the surface excess quantities to the jump in convective and diffusive fluxes from the adjacent regions across the interface.
By specifying $\phi$, $\boldsymbol{j}$, $s$ and their corresponding surface excess quantities it is then possible to obtain the respective transport equations and transmission conditions for mass, momentum energy and entropy. Such specifications under sharp interface assumptions (no interfacial mass) are given in Table \ref{tab:Specifications_of_generic_model}. In this table, $p$ represents the pressure, $\boldsymbol{\tau}$ is the deviatoric stress tensor, $\boldsymbol{g}$ is the gravitational vector, $e$ is the internal energy and $\boldsymbol{q}$ is the heat flux vector. The interfacial stress tensor $\boldsymbol{\tau}^{\Sigma}$ accounts for the capillarity of the interface. The entropy is denoted by $\eta$ with its flux and production in the bulk being represented by $\boldsymbol{j}_{\eta}$ and $\zeta$ respectively. Due to the consideration of interfacial stress, the interfacial entropy production $\zeta^{\Sigma}$ is also accounted for.

\renewcommand{\arraystretch}{1.5}
\begin{table}
\begin{center}
    \caption{Specifications of the general transport equation and the generic interface transmission condition under the assumption of a mass-less but capillary interface}
    \begin{tabular}{l ccc ccc}
        \toprule 
            & $\phi$ & $\boldsymbol{j}$ & $s$ & $\phi^{\Sigma}$ & $\boldsymbol{j}^{\Sigma}$ & $s^{\Sigma}$ \\
        \hline 
            Mass & 1 & 0 & 0 & 0 & 0 & 0   \\
            Momentum & $\boldsymbol{u}$ & $p\mathbf{I}-\boldsymbol{\tau}$ & $\rho\boldsymbol{g}$ & 0 & $-\boldsymbol{\tau}^{\Sigma}$ & 0   \\
            Energy & $e+\frac{\boldsymbol{u}^{2}}{2}$ & $\boldsymbol{q}+ (p\mathbf{I}-\boldsymbol{\tau})\cdot\boldsymbol{u}$ & b & 0 & $-\boldsymbol{\tau}^{\Sigma}\cdot\boldsymbol{u}^{\Sigma}$ & 0   \\
            Entropy & $\eta$  & $\boldsymbol{j}_{\eta}$ & $\zeta$ & 0 & 0 & $\zeta^{\Sigma}$ \\
        \bottomrule
    \end{tabular}
    \label{tab:Specifications_of_generic_model}
\end{center}
\end{table}
\renewcommand{\arraystretch}{1}


\section{Interface-coupling}
\label{sec:interfaceCoupling}
Within the sharp interface model framework, adjacent regions are coupled with each other through the interface transmission condition \eqref{eqn:Local_Interface_Transport}. Note that \eqref{eqn:Local_Interface_Transport} does not represent a boundary condition for the coupling of the primitive fields used to describe the physics of each region. However, it can be reformulated when considering a generic primitive field $f$ \cite{habes_constantin_towards_2023} to match with
\begin{equation} \label{eqn:generic_flux_jump}
      \llbracket \Gamma \nabla f \rrbracket \cdot \boldsymbol{n}_{\Sigma}
    = 
      \mathcal{F}.
\end{equation}
Here, $\Gamma$ typically denotes a constant diffusivity but can also be a dependent function of other variables. The jump of the interfacial flux of $f$ (flux discontinuity) is denoted by $\mathcal{F}$, which might also be a dependent function of other variables. When linearised appropriately, (\ref{eqn:generic_flux_jump}) can be used as a coupling Neumann boundary condition on either side of the interface. 

\noindent
In addition, one also needs to account for jumps of the primitive fields $f$ at the interface in order to fully describe the interfacial coupling. Such jumps of the primitive fields arise from closure, e.g.\ when accounting for the second law of thermodynamics -- see \cite{bothe_sharp-interface_2022} for a rigorous treatise on this subject. Jump conditions can also be written in a generic form, 
\begin{equation} \label{eqn:generic_jump}
      \llbracket f \rrbracket
    = 
      \mathcal{J} \;,
\end{equation}
where $\mathcal{J}$ represents the interfacial jump of $f$ and may be a dependent function of primitive transport variables. In a linearised form (\ref{eqn:generic_jump}) can be applied as a coupling Dirichlet boundary condition on either side of the interface. 

\noindent
When it comes to the algorithmic aspect of region-to-region coupling, monolithic coupling and partitioned coupling types are to be distinguished. In monolithic coupling methods, the coupled equations for each region are assembled and solved simultaneously in one system accounting for the coupling conditions implicitly. Whereas in partitioned coupling methods, the equations of each region are solved separately, updating the coupling conditions using iterations \cite{uekermann_coupling_2014}. These partitioned methods are most often based on so-called non-overlapping domain decomposition methods -- also known as Schwarz methods -- and can be of different types \cite{toselli_domain_2005, smith_domain_2004}. One such type is the Dirichlet-Neumann-Algorithm, which is implemented in the presented framework. Here, each interface is assigned a coupling Dirichlet boundary condition on one side and a coupling Neumann boundary condition on the other side. Since partitioned coupling algorithms in general can lack convergence, acceleration methods are needed during the update procedure \cite{gatzhammer_efficient_2015}. These acceleration methods can either be based on relaxation of the boundary condition values or on quasi-Newton procedures. The \texttt{multiRegionFoam} framework implements a fixed and an Aitken relaxation method as well as the IQN-ILS procedure. The interested reader is referred to  \cite{irons_version_1969, degroote_performance_2010, gatzhammer_efficient_2015} for more details.


\section{Notes on OpenFOAM} \label{sec:notesOnOpenFOAM}

In the following, we attempt to set out details regarding two essential features OpenFOAM provides \emph{by design}, namely the object registry and runtime selection. Both are crucial to understand at this point, since \texttt{multiRegionFoam} leverages them so as to devise a flexible and modular approach for computational multiphysics problems of multi-region coupling type.

\subsection{Run-Time Selection} \label{subsec:RTS}

Typically a domain expert will be concerned with developing models and/or testing different combinations of models. For this purpose, in the strict object-oriented programming paradigm of OpenFOAM, models are implemented as classes answering to the same interface so they are encapsulated and re-usable. Note that the term 'model' is used here in the broadest sense, e.g.\ for the choice of linear solvers, or the selection of discretisation schemes, etc. Then, compiling such model classes into shared objects (model libraries) has the advantage that the selection of models can be deferred until run-time (compared to compile-time in traditional factory methods). Together with Run-Time Selection (RTS) tables, models can then be selected from dynamically loaded shared libraries. Such an approach provides ultimate extensibility, since new models can be just added to a RTS table at run-time (by loading shared libraries dynamically) and become available to the top-level solvers just like the models of the same kind from the legacy code. 
The basic idea of a Run-Time Selection Table is to use a combination of static member variables and methods as well as templates to declare a Hash-Table in the base classes which all child classes register to automatically \cite{noauthor_openfoamwiki_nodate}.
It leverages the fact that when a new shared library is loaded, all static variables are immediately initialized, which is exploited to call code that inserts the type name to the parent class's table of models. The parent class's side basically manages a dynamic 'v-table' to construction methods of child classes. Traditionally, a static method (called New) looks up the requested model (provided by user input), constructs the object with its concrete type, but returns a pointer to the base type. 

\subsection{Object Registry} \label{subsec:OR}

In OpenFOAM the object registry can be thought of as a kind of database that stores information about each object registered to it. It provides a way to keep track of all relevant objects created in a simulation, making it easier to access and manipulate them during runtime. 
An object registry is implemented using a hash table which belongs to C++ data structures that store key-value pairs. In this case, the keys are strings representing the names of the objects, and the values are references to the objects themselves. All classes in OpenFOAM inheriting from {\tt objectRegistry} represent such an object registry. The most notable ones are the classes {\tt Time}, providing read access to e.g.\ mesh objects, and {\tt mesh}, providing read access to e.g.\ field objects. 

\noindent
To look up an object in the registry, OpenFOAM uses a technique called string hashing. If the object is found, its reference is returned. If the object is not found, an exception is thrown indicating that the object does not exist in the registry. In essence, the mechanism relies on two incredients to check for existence of a requested object and return a reference to it (see Listing \ref{lst:objLookup}):
\begin{itemize}
    \item the object's name, which is assumed to be unique in a single database, and
    \item the object's type, which is passed as a template argument to the lookup member function. Dynamic casts are simply used
    to check whether the {\tt regIOobject} object of requested name is also of requested type.
\end{itemize}
\begin{lstlisting}[caption=Object lookup example in OpenFOAM,label={lst:objLookup}]
// Lookup the velocity field (of type volVectorField) from the mesh
const volVectorField& U = mesh.lookupObject<volVectorField>("U");
\end{lstlisting}
\noindent
This hierarchy enables flexible access to objects across multiple libraries. These objects are typically declared at the main scope in solver code and persist throughout the solver's execution. The object registration mechanism facilitates obtaining references to these objects from any shared library, provided there is access to the corresponding database (refer to Section \ref{subsec:RTS} for the significance of this access level). This mechanism effectively eliminates the requirement to pass lengthy lists into class constructors, which otherwise impairs maintainability, extensibility, and generality.

\section{Code structure and design} \label{sec:mutliRegionFoam}

In its essence, \texttt{multiRegionFoam} is a unified framework for multiphysics simulations of region-to-region coupling type. This coverage of distinct region and interfacial physics requires combinatorial flexibility. Therefore, its designed with attention to the following complementary aspects, in many areas going beyond basic requirements of domain-driven software development in research \& development:
\begin{itemize}
    \item \textbf{Usability}. Often domain experts are developing research software using their substantial knowledge on details of the continuum-physical model specific to their own area. To leverage this potential, we have followed the domain-driven software design approach. Utmost attention has been devoted to devise a modular framework for both region- and interface-specific physics which can be developed as entities on their own right. Our aim has indeed been to keep the differences between implementing a module to writing a domain-specific top-level solver in OpenFOAM as low as possible.
    \item \textbf{Understandability}. We have aimed to devise a complete and organized software fabric with a concise, clear as well as descriptive terminology for names of classes, data and functions. This has been motivated by the wish that when presenting \texttt{multiRegionFoam} to an engineer not familiar with the code before, basic functionality and principles of use should be easily comprehended.
    \item \textbf{Generality}. Recognising inherent structural similarities which multiphysics problems of multi-region coupling type have in common, we have devoted significant effort in a general mathematical formulation as foundation of the software design. This has led to a unified framework for multi-region coupling with coverage over a wide range of numerous coupled continuum-physics problems from various distinct fields.
    \item \textbf{Extensibility}. The structure of \texttt{multiRegionFoam} has been purposely designed to allow the flexible and non-intrusive addition of new capabilities or functionality. For instance, it is straight forward to add new modules for region- and interface-specific physics and to complement the set of provided coupling algorithms and coupled boundary conditions if needed.
    \item \textbf{Maintainability}. \texttt{multiRegionFoam} makes comprehensive use of modern C++, such as classes (encapsulation, inheritance and composition), virtual functions (dynamic polymorphism), and operator overloading. We paid attention to enforcing consistent encapsulations of class families under common interfaces being as small as possible. Classes are minimal in size and as low in complexity as possible, so as to fulfill their (single) task. Moreover, special attention has been paid to avoid code-repetition by means of templating.
    \item \textbf{Robustness}. Substantial efforts have been devoted to provide different coupling strategies. The approach enables to deploy both monolithic and partitioned coupling at the user’s choice for each continuum-physical transport equation. This enables to devise the solution strategy to coupled multiphysics problems of region-to-region coupling type on an abstraction level, allowing to have numerical robustness (stability and convergence) in mind despite dealing with substantial model complexity.
    \item \textbf{Efficiency}. We ensure the possibility to efficiently deploy \texttt{multiRegionFoam} on distributed-memory parallel computer architectures on high-performance computing clusters by providing means for straight forward domain-decomposition. In particular, the interface-to-interface communication layer is developed such that it can be used in coupled boundary conditions for data transfer both in serial and parallel in a versatile manner providing multiple mapping strategies.
\end{itemize}
\subsection{Main class structure} 
\label{section:codeStructure}
\begin{figure}[!bh]
    \centering
    \includegraphics[scale=0.8]{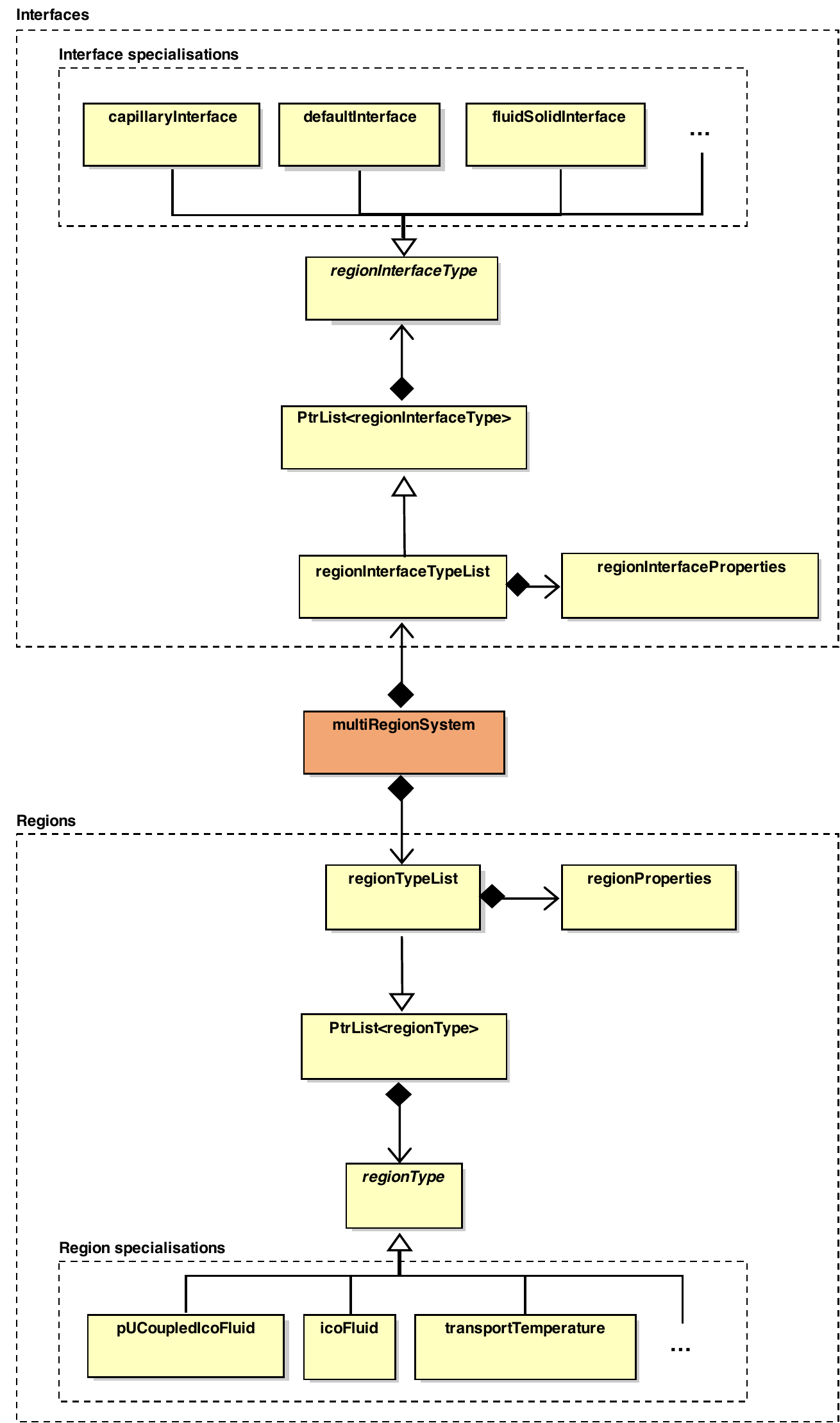}
    \caption{General structure of the \texttt{multiRegionFoam} framework}
    \label{fig:MultiRegionSystem_UML}
\end{figure}

\begin{figure}[H]
    \centering
    \includegraphics[scale=0.4]{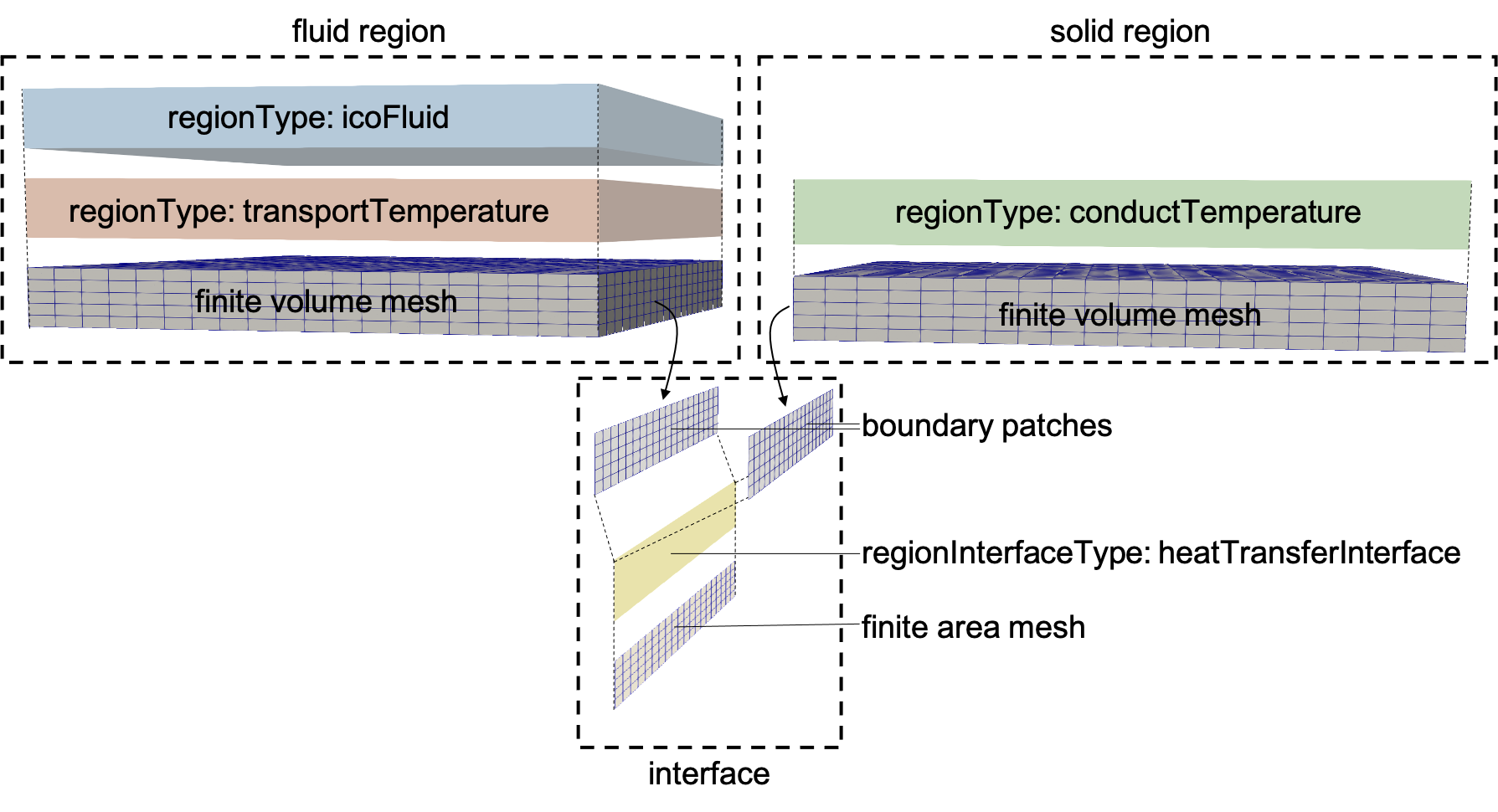}
    \caption{Example of a heat transfer multi region system being composed out of two regions. The physics of the fluid region is described by the superposition of the two \texttt{regionType}s \texttt{icoFluid} and \texttt{transportTemperature} while the physics of the solid region is described by the \texttt{regionType} \texttt{conductTemperature} with a \texttt{heatTransferInterface} coupling both regions}
    \label{fig:MultiRegionSystem_Example}
\end{figure}

We have followed a strictly object-oriented programming paradigm underlying a layered software design. Figure \ref{fig:MultiRegionSystem_UML} depicts the class structure following the unified modeling language (UML) class diagram convention \cite{stevens_using_1999}. 
The building blocks of the code structure are two fundamental classes; \texttt{regionType} and \texttt{regionInterfaceType} which are base classes providing common functionalities that any type of region or interface would require regardless of the simulated physics. Examples of such functionalities for regions include assembling the equations that specify their physical behaviour, correcting the region's material properties and moving its mesh. Similarly for interfaces, examples are administering the protocols for communication and coupling between regions, allowing for deploying various mapping methods, and implementing generic coupled boundary conditions (Section \ref{sec:codeCouplingBCs}).
\noindent
From the aforementioned base classes, a code with modular design is devised which relies on the run-time selection mechanism (Section \ref{subsec:RTS}) in which the fundamental inheritance in C++ plays a crucial role. This enables the creation of derived classes that inherit all common functionalities and extend them to account for additional physical processes at the respective region or interface by defining specialised fields or equations. The inheritance hierarchy is indicated in Figure \ref{fig:MultiRegionSystem_UML} by a solid line with a hollow arrowhead pointing from derived to base classes.  Currently some specialised region types are implemented such as \texttt{icoFluid} for transient flow of incompressible fluid,  \texttt{conductTemperature} and \texttt{transportTemperature} for thermal transport between solid and fluid regions. Also different region interface types are already available, like the \texttt{capillaryInterface} which is a fluid-fluid interface type accounting for surface tension and surfactant transport. Additional region or interface types can be easily added with this design. Moreover, different region types and region interface types can be superimposed. This is made possible by utilising the functionality of the object registry (Section \ref{subsec:OR}) and the helper function \texttt{lookupOrRead} in Listing \ref{lst:objLookupOrRead}, so that all superimposed region types have access to the fields present in one region even if they are defined in other types. 
\begin{lstlisting}[caption=Object lookup or read helper function,label={lst:objLookupOrRead}]
        template <class T>
        inline autoPtr<T> lookupOrRead
        (
            const fvMesh& mesh,
            const word& fldName,
            const bool& read=true,
            const bool& write=true,
            const tmp<T> fld = tmp<T>(nullptr)
        );
\end{lstlisting}
%
The information of the regions and interfaces, such as name, type and the settings for the interfacial coupling algorithm are specified by the user at run-time via the dictionaries \texttt{multi\-RegionProperties} and \texttt{regionInterface\-Properties} (see Listings \ref{lst:regionInterfaceProperties} and \ref{lst:multiRegionProperties} for a simple conjugate heat transfer (CHT) problem whose setup is also displayed in Figure \ref{fig:MultiRegionSystem_Example}). The regions and interfaces information is read and stored by the \texttt{regionProperties} and \texttt{regionInterface\-Properties} classes which are utilised by the \texttt{regionTypeList} and \texttt{regionInterfaceTypeList} classes to instantiate the regions and the interfaces. The \texttt{multiRegionSystem} class builds on these lists to create the multi-region system of equations. It orchestrates the solution process by either applying monolithic or partitioned approaches and therefore acts as the main class interface to the top-level solver \texttt{multiRegionFoam}.
\begin{lstlisting}[caption= \texttt{multiRegionProperties} dictionary for a simple CHT problem,label={lst:multiRegionProperties}]
regions
(
    ( fluid (icoFluid transportTemperature) )
    ( solid (conductTemperature) )
);

DNA // Dirichlet-Neumann Algorithm controls 
{
    T
    {
        maxCoupleIter 20;
        residualControl
        {
            maxJumpRes 1e-07;
            outputJumpResField no;
            maxFluxRes 1e-07;
            outputFluxResField no;
        }
    }
}
\end{lstlisting}
\begin{lstlisting}[caption= \texttt{regionInterfaceProperties} dictionary for a simple CHT problem, label={lst:regionInterfaceProperties}]
partitionedCoupledPatches
(
    fluidsolid
    {
        interfaceType heatTransferInterface;

        coupledPatchPair
        (
            (fluid bottom)
            (solid top)
        );

        coupledFields
        (
            T
        );
        
        defaultInterfaceCoeffs{}
    }
);
monolithicCoupledPatches
(
);
curvatureCorrectedSurfacePatches 0();
interpolatorUpdateFrequency 1;
interfaceTransferMethod directMap;
directMapCoeffs{}
GGICoeffs{}
\end{lstlisting}
\subsection{Solution of the coupled system} 
\label{section:codeListings}
The set of equations describing the physical behaviour are defined in the specialisations of the \texttt{regionType} class in the \texttt{setCoupledEqns} function as illustrated in Listing \ref{lst:regionTypes::transportTemperature::setCoupledEqns} for the transport temperature equation in a fluid region. An instance of the equation system is stored in a hash pointer table, \texttt{HashPtrTable}. In this case, it is called \texttt{fvScalarMatrices} which points to a finite volume matrix system of scalar type. Other types include vector, tensor, or symmetric tensor matrices, as well as block coupled types \texttt{fvBlockMatrix}, as indicated in Listing \ref{lst:regionTypeH_coupled governing equations}. A unique name identifier is used as a key for the hash table where the same name pattern is used among all types of regions for straightforward access to all coupled equations later when the system is set up and solved. 
%
\begin{lstlisting}[caption=\texttt{setCoupledEqns} of \texttt{transportTemperature} region type,  label={lst:regionTypes::transportTemperature::setCoupledEqns}]
void Foam::regionTypes::transportTemperature::setCoupledEqns()
{
    //- Create temperature equation system
    TEqn =
    (
        rho_*cp_
       *(
            fvm::ddt(T())
          + fvm::div(phi_(), T())
        )
     ==
        fvm::laplacian(kappa_(), T())
    );
    //- Store equation system in appropriate HashPtrTable 
    fvScalarMatrices.set
    (
        T_().name()
      + mesh().name() + "Mesh"
      + transportTemperature::typeName + "Type"
      + "Eqn",
        &TEqn()
    );
}
\end{lstlisting}
\begin{lstlisting}[showstringspaces=false, caption=coupled governing equations, label={lst:regionTypeH_coupled governing equations}]
//- coupled governing equations
HashPtrTable<fvMatrix<scalar> > fvScalarMatrices;
HashPtrTable<fvMatrix<vector> > fvVectorMatrices;
HashPtrTable<fvMatrix<symmTensor> > fvSymmTensorMatrices;
HashPtrTable<fvMatrix<tensor> > fvTensorMatrices;
HashPtrTable<fvBlockMatrix<vector4> > fvVector4Matrices;
\end{lstlisting}
The multi-region system comprising all such physics-specific equations, defined in different specialisations of region types, is assembled and solved in the  \texttt{multiRegionSystem} class via the \texttt{solve} function shown in Listing \ref{lst:multiRegionSystem::solve}. The auto pointers \texttt{interfaces\_} and \texttt{regions\_} which point to \texttt{regionInterfaceTypeList} and \texttt{regionTypeList} classes (cf.\ Figure \ref{fig:MultiRegionSystem_UML}) provide access to the list of all regions and their interfaces, including the functions defined in their respective specialised classes. For example, using \texttt{interfaces\_->detach()}, the detach boundary mesh modifier is applied to the interfaces in preparation for partitioned coupling while \texttt{regions\_->solveRegion()} checks for possible individual physics solution requirement in each region. Another function, \texttt{solvePIMPLE} (see Section \ref{sec:code:solvePimple}), is dedicated for solving pressure-velocity systems, if any, using the PIMPLE algorithm. A distinction of cases is made based on whether the pressure and velocity fields are coupled across the interface or not. For example, solving the pressure-velocity system in the fluid region in the case of heat transfer between fluid and solid, versus the case of pressure-velocity interfacial coupling between two fluids as in a rising bubble scenario. Moreover, the coupling of the pressure and the velocity fields is made separate from other fields in order to allow for mesh motion using the Arbitrary Lagrangian-Eulerian (ALE) interface tracking method. For other fields, the system of equations is created and solved under the \texttt{assembleAndSolveEqns} function for partitioned coupling and the \texttt{assembleAndSolveCoupledMatrix} for monolithic coupling. Each of these functions loops over the list of the partitioned or monolithic coupled fields from the lists \texttt{partitionedCoupledFldNames} or \texttt{monolithicCoupledFldNames}, respectively. These are \texttt{hashedWordList}s reading coupled fields as specified in the \texttt{regionInterfaceProperties} dictionary (Listing \ref{lst:regionInterfaceProperties}). Note that when the partitioned Dirichlet-Neumann Algorithm (DNA) (Section \ref{sec:interfaceCoupling}) is selected, a convergence criteria is added. It is based on the interface residuals (see Section 4.4.2 in \cite{habes_constantin_towards_2023}) which are computed in the generic coupled boundary condition classes (Section \ref{sec:codeCouplingBCs}). \texttt{dnaControl} has access to these coupled boundary conditions and residuals for all interfaces. The DNA controlled fields and the termination criteria are specified by the user in the \texttt{multiRegionProperties} dictionary as illustrated for the temperature field \texttt{T} in Listing \ref{lst:multiRegionProperties}  under the \texttt{DNA} entry.
\begin{lstlisting}[showstringspaces=false, caption=\texttt{solve} function of \texttt{multiRegionSystem}, label={lst:multiRegionSystem::solve}]
void Foam::multiRegionSystem::solve()
{
    //- Detach boundary mesh modifier
    interfaces_->detach();
    
    //- Solve individual region physics
    regions_->solveRegion();
    
    //- Check if at least one region implements PIMPLE loop
    //  and solve pressure-velocity system if so
    if (regions_->usesPimple())
    {
        // PIMPLE p-U-coupling
        regions_->solvePIMPLE();
    }
    
    //- Solve region-region coupling (partitioned)  
    forAll (partitionedCoupledFldNames_, fldI)
    {
        //- Get name of field which is
        // currently partitioned coupled 
        word fldName = partitionedCoupledFldNames_[fldI];

        //- Solve pressure-velocity system using PIMPLE
        if (fldName == "pUPimple")
        {
            while (dnaControls_[fldName]->loop())
            {
                // PIMPLE p-U-coupling
                regions_->solvePIMPLE();
                // ALE mesh motion corrector
                regions_->meshMotionCorrector();
                // Update interface inherent physics
                interfaces_->update();
            }
        }
        else
        {
            //- Solve other partitioned coupled fields
            while (dnaControls_[fldName]->loop())
            {
                assembleAndSolveEqns<fvMatrix, scalar>(fldName);
                // fields of higher tensor rank
                // ... 
            }
        }
    }
    //- Attach boundary mesh modifier
    interfaces_->attach();

    //- Solve region-region coupling (monolithic)
    forAll (monolithicCoupledFldNames_, fldI)
    {
        //- Get name of field which is 
        // currently monolithic coupled 
        word fldName = monolithicCoupledFldNames_[fldI];

        assembleAndSolveCoupledMatrix<fvMatrix, scalar>
        (
            monolithicCoupledScalarFlds_, fldName
        );
        // fields of higher tensor rank
        // ... 
    }
}
\end{lstlisting}
The assembly and solution of the system of coupled equations in partitioned mode is shown in Listing \ref{lst:multiRegionSystem::assembleAndSolveCoupledEqns} which represents the \texttt{assembleAndSolveEqns} function of the \texttt{multiRegionSystem} class. The function iterates through all regions and obtains a non-constant reference, called \texttt{rg}, to the current region. It also constructs a unique name \texttt{matrixSystemName} for the equation system for the coupled field in this region. This name matches the pattern of the key assigned for the hash pointer table defined within the \texttt{setCoupledEqns} function in the respective specialised region type as mentioned earlier (see Listing  \ref{lst:regionTypes::transportTemperature::setCoupledEqns}). The coupled equation matrix defined in the current region is retrieved from the hashed table using the \texttt{getCoupledEqn} function which is a member of the \texttt{regionType} class. It is defined using a template, which allows it to be specialised for different types of matrices where the type is deduced from the input \texttt{matrixSystemName}. The use of templatisation is particularly beneficial here to avoid writing multiple identical classes differing only in type. The system of equations is then assembled and solved with the option to perform individual post-solve actions that are implemented in each region.
\begin{lstlisting}[showstringspaces=false, caption=\texttt{assembleAndSolveEqns} of \texttt{multiRegionSystem}, label={lst:multiRegionSystem::assembleAndSolveCoupledEqns}]
template< template<class> class M, class T>
void Foam::multiRegionSystem::assembleAndSolveEqns
(
    word fldName
) const
{
    forAll (regions_(), regI)
    {
        //- Getting non const access to current region
        regionType& rg = const_cast<regionType&>(regions_()[regI]);
        
        //- Unique name of equation system
        word matrixSystemName =
        (
            fldName
          + rg.mesh().name() + "Mesh"
          + rg.regionTypeName() + "Type"
          + "Eqn"
        );
        
        //- Sanity checks
        // ...
        
        //- Get equation from region 
        M<T>& eqn = rg.getCoupledEqn<M,T>(matrixSystemName);
        
        //- Relax equation
        rg.relaxEqn<T>(eqn);
        
        //- Solve equation
        eqn.solve();
        
        //- Post solve actions
        rg.postSolve();
    }
}
\end{lstlisting}
For monolithic coupling, the procedure starts with attaching the meshes at the interfaces between adjacent regions using the attach boundary mesh modifier \texttt{interfaces\_->attach()} function as indicated previously in Listing \ref{lst:multiRegionSystem::solve}. Then, as for partitioned coupling, the \texttt{assembleAndSolveCoupledMatrix} function creates and solves the coupled system. However, it is now represented as a block coupled finite volume matrix using the \texttt{coupledFvMatrix} approach \cite{noauthor_coupledFvMatrix_nodate} as illustrated in Listing \ref{lst:multiRegionSystem::assembleAndSolveCoupledMatrix}. This is a block matrix system that is initialised with the number of monolithic coupled regions. The equations to be loaded into the \texttt{coupledFvMatrix} are obtained from the respective regions using the \texttt{getCoupledEqn} function analogously to the partitioned approach by iterating over the regions and collecting equations by their unique name. Unlike partitioned coupling, the retrieved coupled equations are not directly solved but instead inserted into the block matrix system. They are first appended into the dynamic list of equations \texttt{eqns} which holds pointers to the equation matrices for all regions. The \texttt{new} operator dynamically allocates memory for each instance of a coupled equation appended to the list. Finally, the equations are solved in the block matrix system simultaneously in an implicit manner using the solver specified by the solution dictionary of the first region's mesh for the given coupled field name.
\begin{lstlisting}[showstringspaces=false, caption=\texttt{assembleAndSolveCoupledMatrix} of \texttt{multiRegionSystem}, label={lst:multiRegionSystem::assembleAndSolveCoupledMatrix}]
template< template<class> class M, class T>
void Foam::multiRegionSystem::assembleAndSolveCoupledMatrix
(
    PtrList<GeometricField<T, fvPatchField, volMesh> >& flds,
    word fldName
) const
{
    //- Get number of monolitic coupled regions per field name
    //  and return if there are none
    // ...

    //- Initialise block matrix system 
    //  with number of monolitic coupled regions
    coupledFvMatrix<T> coupledEqns(nEqns);

    //- Assemble all matrices one-by-one and combine them into the
    //  block matrix system
    label nReg = 0;
    DynamicList<M<T>* > eqns;
    forAll (regions_(), regI)
    {
        //- Getting non const access to current region
        regionType& rg = const_cast<regionType&>(regions_()[regI]);

        //- Unique name of equation system
        word matrixSystemName =
        (
            fldName
          + rg.mesh().name() + "Mesh"
          + rg.regionTypeName() + "Type"
          + "Eqn"
        );

        //- Get equation from region 
        M<T>& eqn = rg.getCoupledEqn<M,T>(matrixSystemName);

        //- Insert matrix into block matrix system
        eqns.append(new M<T>(eqn));
        coupledEqns.set(nReg, eqns[nReg]);
        nReg++;
    }

    //- Solve block matrix system
    coupledEqns.solve
    (
        regions_()[0].mesh().solutionDict().solver(fldName + "coupled")
    );

    // Post solve actions for monolithic coupled fields
    // ...
}
\end{lstlisting}
\subsection{Pressure-velocity coupling}  \label{sec:code:solvePimple}
The pressure-velocity coupling is solved in a semi-implicit manner using the PIMPLE algorithm which is a combination of the well known Pressure-Implicit with Splitting of Operators (PISO) algorithm \cite{issa_solution_1986} and a more consistent version of the Semi-Implicit Method for Pressure-Linked Equations (SIMPLE) algorithm \cite{patankar_calculation_1972}. This involves a predictor-corrector procedure in which a tentative solution for the velocity is obtained in the predictor step while the pressure is updated in the corrector step. The standard implementations of these algorithms in OpenFOAM are refactored and integrated into the class structure of \texttt{multiRegionFoam} (see Figure \ref{fig:MultiRegionSystem_UML}). The code is implemented in the \texttt{icoFluid} region type class which needs to be specified by the user in the \texttt{multi\-RegionProperties} dictionary (Listing \ref{lst:multiRegionProperties}). The code block for each of the steps of the PIMPLE algorithm are defined in the \texttt{momentumPredictor} and the \texttt{pressureCorrector} functions. These are called in the \texttt{regionTypeList} class by utilising the concept of operator overloading in C++, as illustrated in Listing \ref{lst:regionTypeList::solvePIMPLE}. The \texttt{solvePIMPLE()} function represents the skeleton of the PIMPLE algorithm which outlines the outer and the inner loops and the execution across regions. In order to ensure the consistency of the pressure-velocity coupling, each step of the solution procedure is performed for all regions before the next step is carried out. 
\begin{lstlisting}[showstringspaces=false, caption=\texttt{solvePIMPLE} of \texttt{regionTypeList},label={lst:regionTypeList::solvePIMPLE}]
void Foam::regionTypeList::solvePIMPLE()
{
    //- Get the number of outer correctors that should be performed  
    //  during the PIMPLE procedure (We do not have a top-level 
    //  mesh. Construct fvSolution for the runTime instead.)
    fvSolution solutionDict(runTime_);
    const dictionary& pimple = solutionDict.subDict("PIMPLE");
    int nOuterCorr(readInt(pimple.lookup("nOuterCorrectors")));

    //- PIMPLE loop
    for (int oCorr=0; oCorr<nOuterCorr; oCorr++)
    {
        forAll(*this, i)
        {
            this->operator[](i).prePredictor();
        }

        forAll(*this, i)
        {
            this->operator[](i).momentumPredictor();
        }

        forAll(*this, i)
        {
            this->operator[](i).pressureCorrector();
        }
    }
}
\end{lstlisting}

\subsection{Generic boundary conditions for partitioned coupling} 
\label{sec:codeCouplingBCs}
\begin{figure}[H]
    \centering
    \includegraphics[scale=0.91]{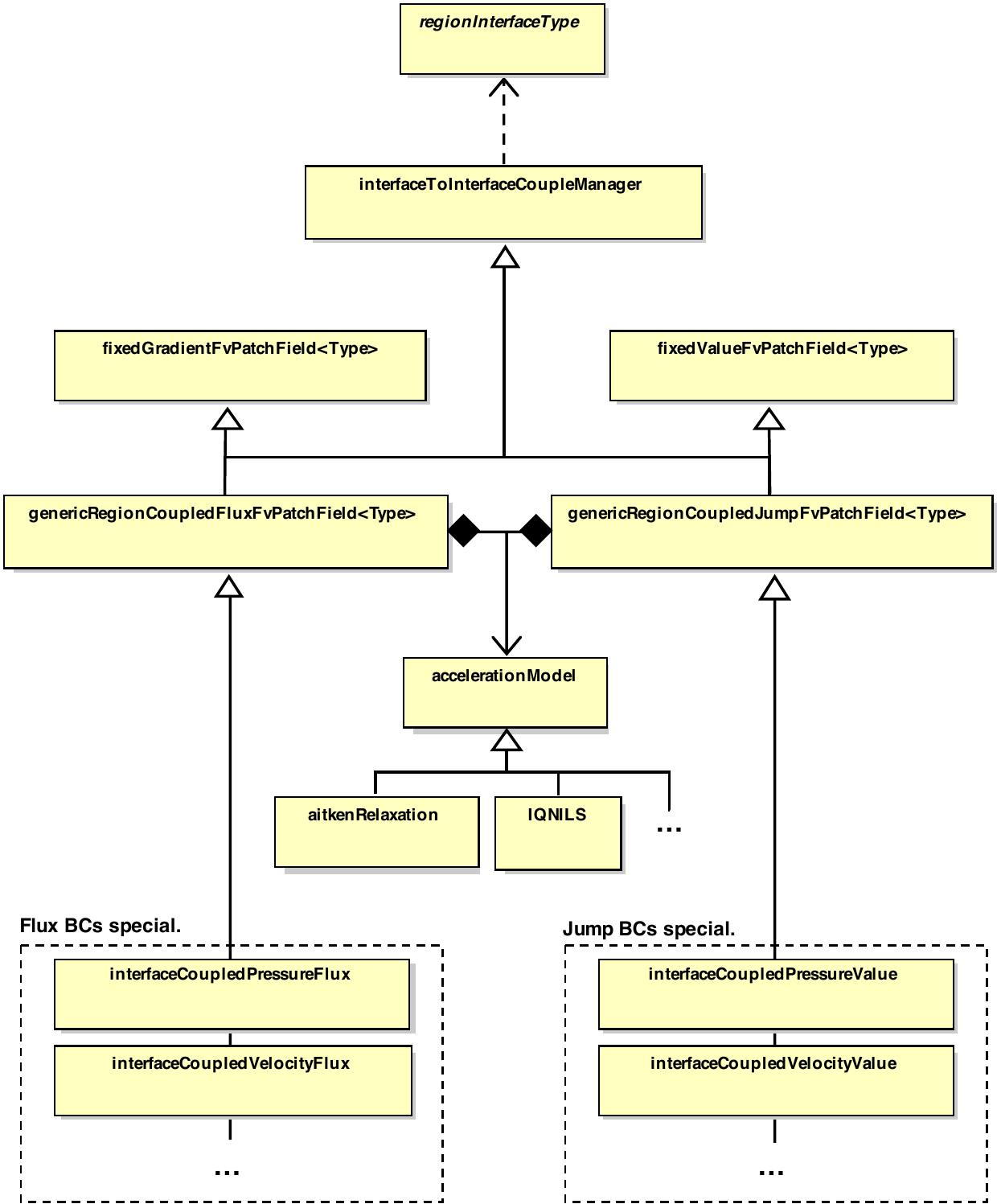}
    \caption{General structure of the generic coupled boundary condition}
    \label{fig:GenericCoupledBCs_UML}
\end{figure}

As described in Section \ref{sec:interfaceCoupling}, the solution of the multi-region system using a partitioned approach requires specifying boundary conditions on the connecting boundaries of the sub-regions where a Dirichlet condition is applied on one side while a Neumann condition is applied on the other one. For this purpose, a set of generic boundary conditions are devised exploiting the fact that thes conditions have a general mathematical formulation (Section \ref{sec:mathModel}). Figure \ref{fig:GenericCoupledBCs_UML} shows how the implementation is integrated into the structure of \texttt{multiRegionFoam}. 
\noindent
The generic boundary conditions are derived from the standard OpenFOAM Dirichlet and Neumann boundary conditions, namely \texttt{fixedValueFvPatchField} and \texttt{fixedGradientFvPatchField}. This results in two classes \texttt{genericRegionCoupledValueFvPatchField} and \texttt{genericRegionCoupledFluxFvPatchField} which implement the interfacial jump and transmission conditions, respectively. Utilising templates, these conditions are made agnostic to the actual physics that the jumps and fluxes originate from and rather receive this information from the relevant specialisation of \texttt{regionInterfaceType}s. The derived coupled boundary conditions also inherit from the \texttt{interfaceToInterfaceCoupleManager} class which gives access to the neighbour region data, such as the mesh and patch names, and has access to the \texttt{regionInterfaceType} on which the coupled boundary condition is applied. For their use in the partitioned interface coupling solution approach they are also able to utilize different convergence acceleration methods including, besides others, the Aitken relaxation method \cite{gatzhammer_efficient_2015} and the Interface Quasi-Newton Inverse Least-Squares method (IQN-ILS) \cite{davis_enhancing_2022}.

\subsection{Parallelisation}
\begin{figure}[H]
    \centering
    \includegraphics[scale=0.6]{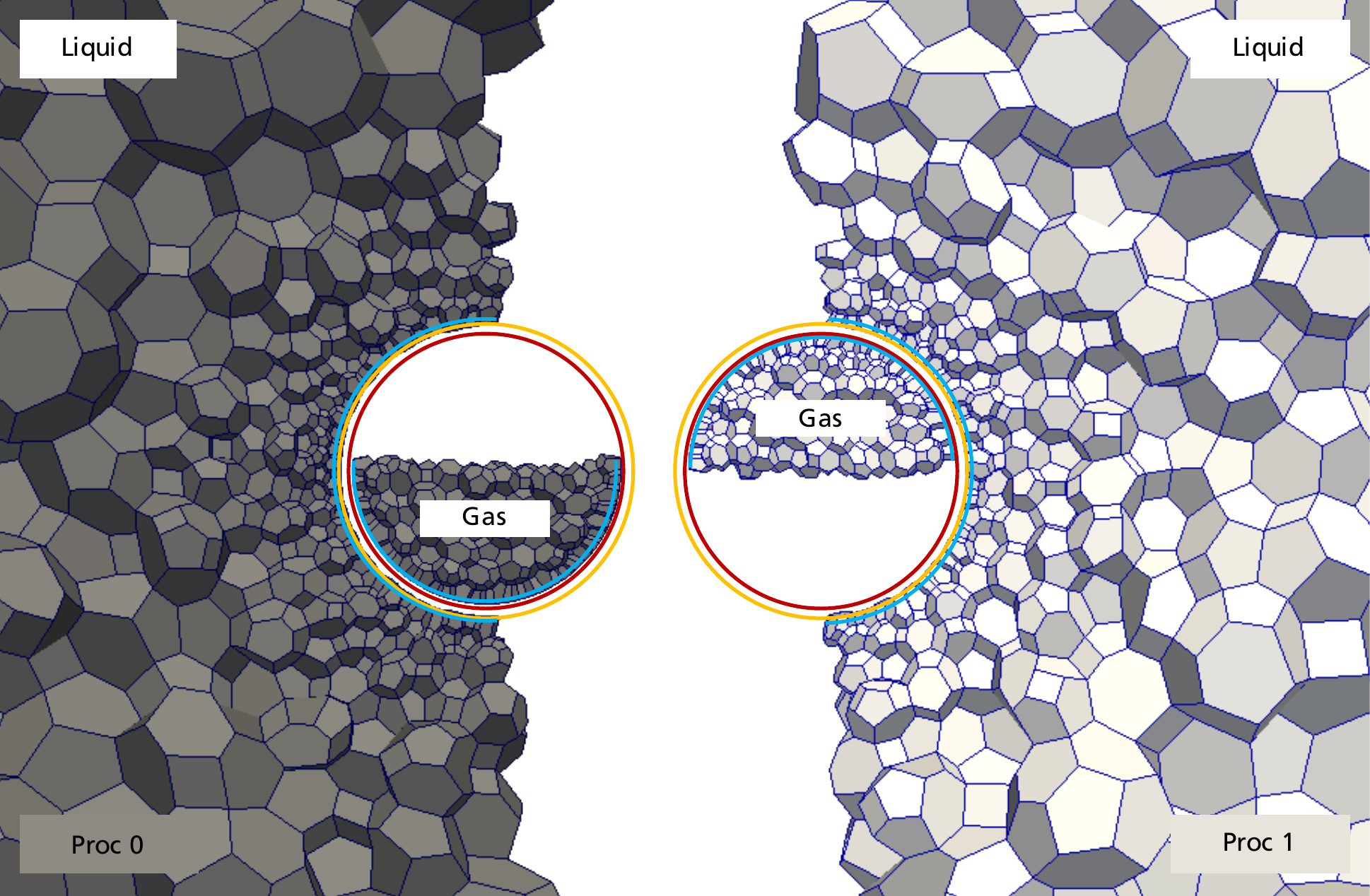}
    \caption{Local (blue) and global (red, orange) poly patches of a gas and a liquid region decomposition}
    \label{fig:GlobalPolyPatches}
\end{figure}

In OpenFOAM, parallelization is generally implemented through the domain decomposition method. This typically requires dividing the solution domain into sub-regions in order for each of them to be solved on separate CPU cores. The standard Message Passing Interface (MPI) \cite{gropp_high-performance_1996} is then utilised to establish communication between the processors. Although the proposed multi-region framework already requires subdivision of the computational mesh into multiple domains, these sub-domains are not convenient for efficient parallel computing, especially because a unified framework is sought. For example, if partitioned coupling is selected, then only one region will be solved at a time, or if the regions vary significantly in size in some scenarios, this will lead to an unbalanced distribution of load over the processors. Instead, all regions are decomposed into the same number of sub-domains but not necessarily using the same decomposition method. Figure \ref{fig:GlobalPolyPatches} depicts an example of two interface coupled regions representing gas bubble and liquid where both phases are decomposed into two sub-domains but one is decomposed vertically while the other is decomposed horizontally. This causes the two patches, that form the two sides of the interface, to be non-entirely overlapping (cf.\ blue lines in Figure \ref{fig:GlobalPolyPatches}) which makes the fields mapping between the two sides impossible. To overcome this situation, the \textit{global face zone} approach, introduced by Cardiff et al. \cite{cardiff_open-source_2018} and Tuković et al. \cite{tukovic_openfoam_2018}, is used. It suggests that each processor is given access to the entire interface patch via the so-called \texttt{globalPolyPatch} (red and orange lines in Figure \ref{fig:GlobalPolyPatches}) which resembles the union of the non-overlapping boundaries, holding copies of their data and allowing for mapping the fields from one global patch to the other. These global patches also facilitate the implementation of the coupled boundary conditions. To adapt this concept in \texttt{multiRegionFoam}, the \texttt{globalPolyPatch} is incorporated into the \texttt{regionInterfaceType} class, i.e.\ it includes a pair of the local interface patches as well as pair of global patches.       
%
\section{Automated test harness} 
In order to validate and evaluate different aspects of \texttt{multiRegionFoam}, an automated test framework is deployed. The main goal is to automatically run a suite of test cases and report on the results. This includes testing whether each simulation was successfully executed and completed as well as checking the obtained results against some criteria, such as a reference solution or an error threshold. The testing framework is also useful for parameter studies, continuous integration, and testing of further developments or new features. The tests are performed using Python utilising the \texttt{oftest} library \cite{noauthor_oftest_nodate} which is a test framework for OpenFOAM cases available as open source under the GPLv3 License. It makes use of the \texttt{pytest} library such that, by adding the script file \texttt{test\_*.py} to each test case, \texttt{pytest} automatically discovers all tests in the folder. This script file includes all the instructions to initiate the simulation and store the results for each test case. The \texttt{run\_reset\_case} module from \texttt{oftest} library is responsible for running and cleaning the case according to the standard OpenFOAM scripts \texttt{Allrun} and \texttt{Allclean}. Listing \ref{lst:TestHarnessRunComplete} shows the \texttt{test\_completed} function which adds the case to the pool of tests, returns the status of completion (passed/failed) in the terminal, and leaves the simulation log file in the specified path. 
\begin{lstlisting}[showstringspaces=false, language=Python, caption= Test case completion check using \texttt{oftest} and \texttt{pytest} libraries,label={lst:TestHarnessRunComplete}]
def test_completed(run_reset_case):
        log = oftest.path_log() 
        assert oftest.case_status(log) == 'completed' 
\end{lstlisting}
For parameter studies and/or testing different scripts for the same case, the \texttt{oftest.Case\_modifiers} module is used allowing for the manipulation of the case files and dictionaries. For example, Listing \ref{lst:testHarnessSetup} shows part of the script file for a suite of tests for a conjugate heat transfer scenario where different values of parameters are simulated, including Reynolds number, Re, the Prandtl number, Pr, and the thermal conductivity ratio, $\operatorname{k}= k_s/k_f$ (see Section \ref{sec:FOHP} for the flow over heated plate case description). Additionally, the test includes different coupling algorithm, monolithic (\texttt{"Allrun -m"}) or partitioned (\texttt{"Allrun -p"}), where the latter could be performed with different types of acceleration \texttt{accType} (\texttt{"aitken"}/ \texttt{"IQN-ILS"}) or without acceleration (\texttt{"fixed"}). In order to execute the tests, the \texttt{@pytest.mark.parametrize} decorator is used to specify the test parameters and their corresponding values as shown in Listing \ref{lst:testHarnessMarkParametrize} which is included in the same script file. 
\begin{lstlisting}[showstringspaces=false, language=Python, caption=  Test suites setup in the script file \texttt{test\_*.py} ,label={lst:testHarnessSetup}]
def paramSet(Pr, Re, k, relaxType, relaxValue, script):
    L = 1.0
    rhof = 1.0
    ks = 100.0
    Uinf = 1.0
    mu = rhof*Uinf*L/int(Re)
    kf = ks/k
    cp = kf*Pr/mu
    
    dir_name = os.path.dirname(os.path.abspath(__file__))
    filemod = 
    {
        "constant/fluid/transportProperties": 
        [
            ('mu', 'mu [ 1 -1 -1 0 0 0 0 ] %s' %mu),
            ('cp', 'cp [ 0 2 -2 -1 0 0 0 ] %s' %cp),
            ('k', 'k [1 1 -3 -1 0 0 0] %s' %kf),
        ],
        "0/fluid/orig/monolithic/k": 
        [
            ('internalField', 'uniform %s' %kf),
        ],
        "0/fluid/orig/partitioned/T": 
        [
            ('boundaryField/interface/accType', relaxType),
            ('boundaryField/interface/relax', relaxValue),
        ]
    }
    meta_data = {"Pr":Pr, "Re": Re, "k": k, "relaxType": relaxType, 
    "relaxValue":relaxValue, "script": script}
    case_mod = oftest.Case_modifiers(filemod, dir_name, meta_data)
    return case_mod
\end{lstlisting}
%
%
During the test execution, \texttt{pytest} will run the \texttt{test\_flowOverHeatedPlate} function multiple times, once for each set of input values specified in the decorator's parameters list. Before cleaning the current case and proceeding to the next one, the data of interest from the current case, such as solution fields at final time, outcomes of OpenFOAM post-processing utilities, or computed errors are stored and written into a file format supported by Python as demonstrated in Listing \ref{lst:testHarnessMarkParametrize}. The results of all tests could be written into a single data frame file that can be further analyzed using standard Python libraries for data management and visualisation. 

\begin{lstlisting}[showstringspaces=false, language=Python, caption= Test suites execution in the script file \texttt{test\_*.py} ,label={lst:testHarnessMarkParametrize}]

parameters = [paramSet(0.01, 500, 1, "fixed", 1, "Allrun -p"), 
              paramSet(0.01, 500, 1, "aitken", 0.75, "Allrun -p"),
              paramSet(0.01, 500, 1, "monolithic", 0, "Allrun -m"),
              paramSet(100, 500, 1, "fixed", 1, "Allrun -p"),
              ... ]

labels = ["Pr01Re500k1Fixed1", "Pr01Re500k1Aitken75", 
          "Pr01Re500k1Monolithic", "Pr100Re500k1Fixed1", ... ]

@pytest.mark.parametrize(
    "run_reset_case", parameters,
    indirect=["run_reset_case"], ids = labels)

def test_flowOverHeatedPlate(run_reset_case, load_results):

    # Access the current case parameters from the meta_data dictionary
    current_case = run_reset_case.meta_data
    
    # Append the error retrieved by load_results function 
    current_case['error'] = max(load_results['error'])
    
    # Append current case to the list of cases 
    all_cases.append(current_case)

    # Write results into data frame in csv format
    all_cases_df = pd.DataFrame(all_cases)
    all_cases_df.to_csv(os.path.join(dir_name, "all_cases.csv"),index=False)

    # Report on test completion
    log = oftest.path_log()    
    assert oftest.case_status(log) == "completed"
\end{lstlisting}


\newpage
\section{Examples of Usage}\label{sec:usage}

\subsection{Forced convection heat transfer from a flat plate}
\label{sec:FOHP}

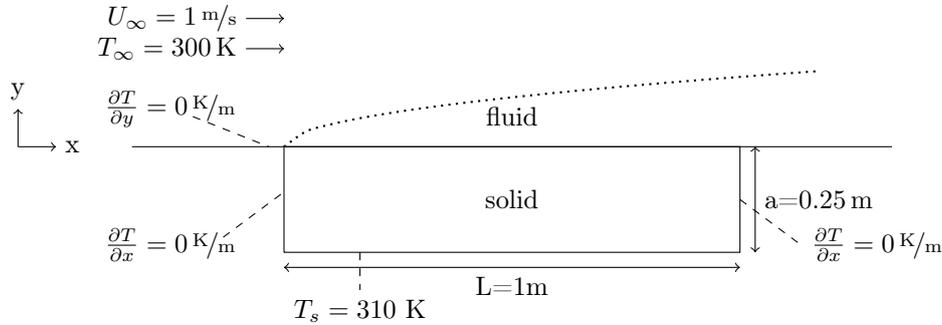
\begin{figure}[H]
\centering
\begin{tikzpicture}

\draw[-] (0,1.4) -- (10,1.4);

\draw[->] (-1.5,1.4) -- (-1,1.4) node[right] {x};

\draw[->] (-1.5,1.4) -- (-1.5,1.9) node[above] {y};

\coordinate (A) at (2,1.5);
\coordinate (B) at (10,1.7);
\draw[domain=0:1,smooth, thick, dotted,variable=\x] plot (7*\x+2,{sqrt(\x)+1.4});

\node at (5,1.8) {fluid};

\draw (2,0) -- (8,0) -- (8,1.4) -- (2,1.4) -- (2,0);
\node at (5,0.7) {solid};

\draw[<->] (8.2,0) -- (8.2,1.4) node[midway, right] {a=$\unit[0.25]{m}$};

\draw[dashed] (8,0.7) -- (8.8,0.1);
\node at (9.8,0.1) {$\frac{\partial T}{\partial x} = \unitfrac[0]{K}{m} $};

\draw[<->] (2,-0.2) -- (8,-0.2) node[midway, below] {L=1m};

\draw[dashed] (3,0) -- (3,-0.5) node [below] {$T_s = 310$ K};

\draw[dashed] (1.2,0.2) -- (2,0.8);
\node at (0.5,0.1) {$\frac{\partial T}{\partial x} = \unitfrac[0]{K}{m} $};

\draw[dashed] (0.8,1.8) -- (1.8,1.4);
\node at (0.5,1.9) {$\frac{\partial T}{\partial y} = \unitfrac[0]{K}{m} $};

\draw[->] (1.5,3.1) node [left] {$U_{\infty} = \unitfrac[1]{m}{s}$} --  (2,3.1);
\draw[->] (1.5,2.7) node [left] {$T_{\infty} = \unit[300]{K}$} --  (2,2.7);

\end{tikzpicture}
\caption{Computational domain and boundary conditions for the flow over a heated plate}
\label{fig:FOHPDomain}
\end{figure}

\begin{figure}[htp]
\centering
\subfloat[Dimensionless conjugate boundary temperature $\theta$ over non-dimensional distance $x$ for Pr = 0.01 and Re = 500]{%
  \includegraphics[clip,width=0.85\columnwidth, scale=0.03]{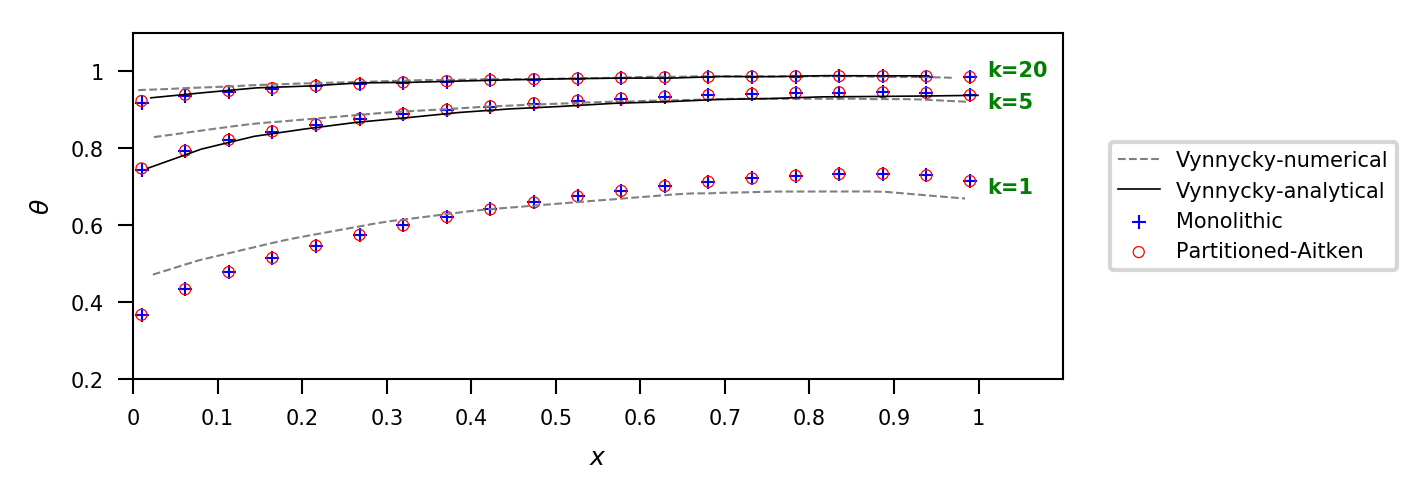}%
}
\vspace{-0.1\baselineskip}
\subfloat[Dimensionless conjugate boundary temperature $\theta$ over non-dimensional distance $x$ for Pr = 0.01 and Re = $10^4$]{%
  \includegraphics[clip,width=0.85\columnwidth, scale=0.03]{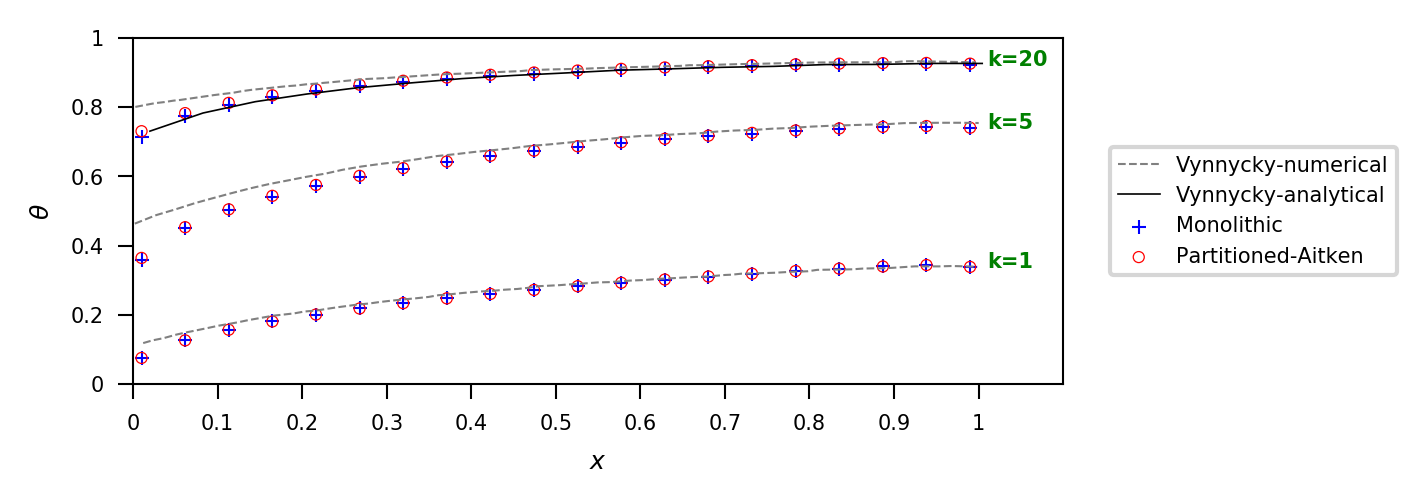}%
}
\vspace{-0.1\baselineskip}
\subfloat[Dimensionless conjugate boundary temperature $\theta$ over non-dimensional distance $x$ for Pr = 100 and Re = 500]{%
  \includegraphics[clip,width=0.85\columnwidth, scale=0.03]{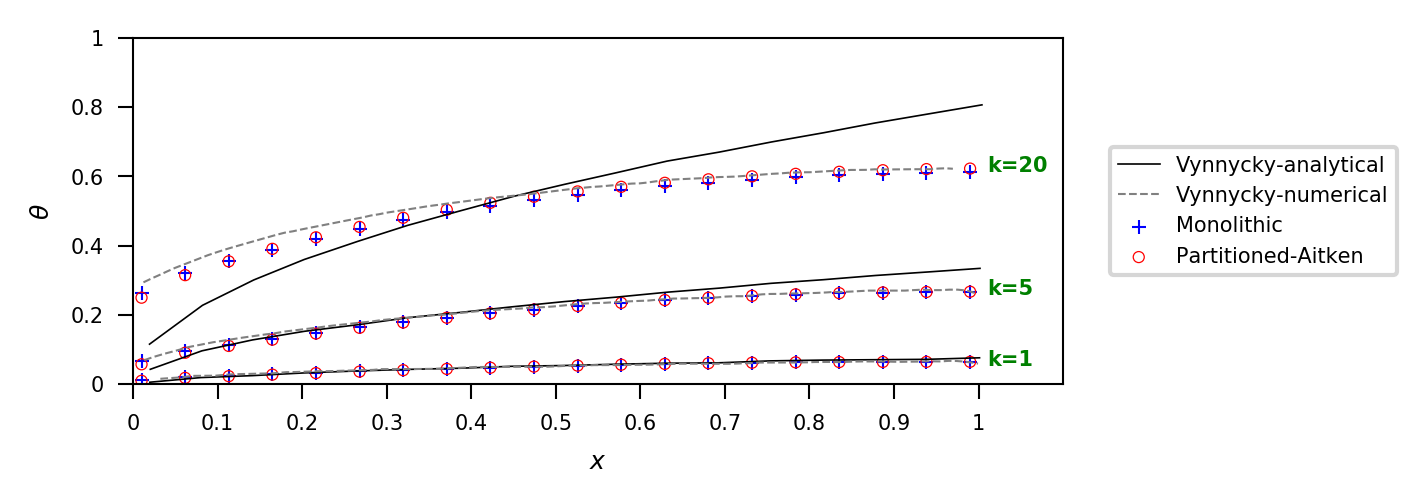}%
  \label{fig:FOHPResultsPr100Re500}
}
\vspace{-0.1\baselineskip}
\subfloat[Average Coupling time for partitioned and monolithic coupling]{%
  \includegraphics[clip,width=0.9\columnwidth, scale=0.025]{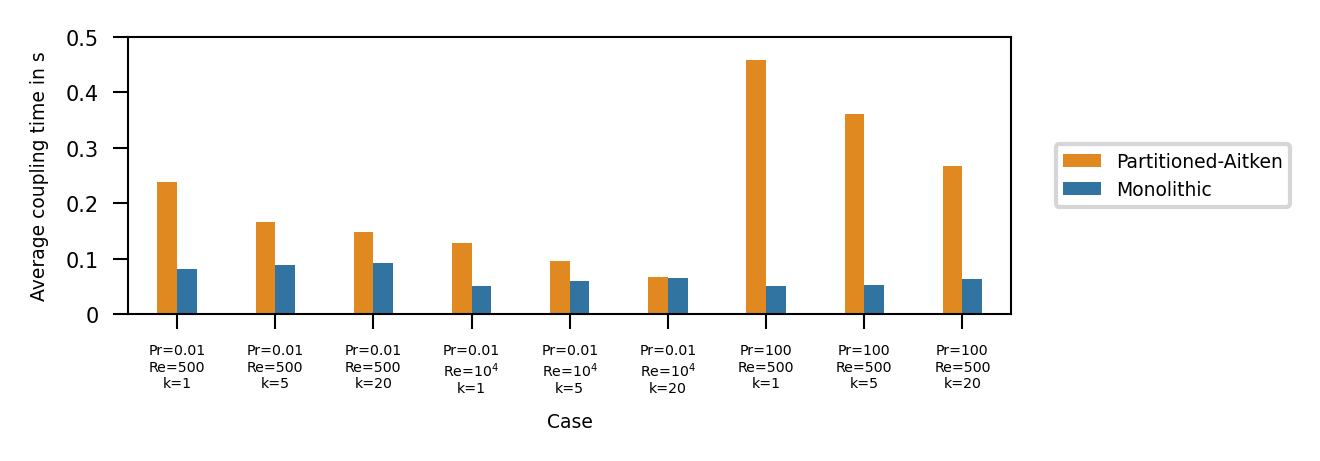}%
  \label{fig:FOHPResultsAvgCouplingTime}
}
\caption{Simulation results for different $Pr$, $Re$, and $k$ values}
\label{fig:FOHPResults}
\end{figure}

In this case, an incompressible laminar flow over a flat plate is considered, following the description from Vynnycky et al. \cite{vynnycky_forced_1998}. Their numerical results as well as their derived reference solution, using boundary-layer theory, are used for validation. A schematic sketch of the case setup is shown in Figure \ref{fig:FOHPDomain}. A fluid of uniform temperature $T_{\infty}$ and velocity $U_{\infty}$ flows over a plate of finite thickness that is held at constant temperature $T_s > T_{\infty}$. 
\begin{figure}[H]
    \centering
    \includegraphics[width=\textwidth]{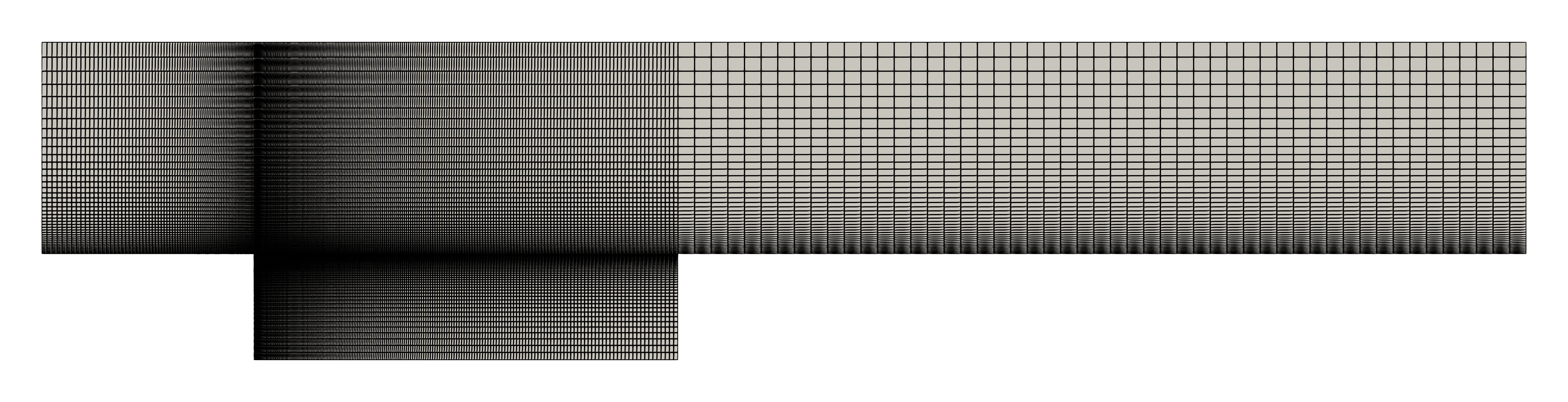}
    \caption{Meshes for the flow over heated plate simulation}
    \label{fig:FOHPmesh}
\end{figure}
\noindent
The computational mesh is depicted in Figure \ref{fig:FOHPmesh}. It covers the fluid and solid regions, both consisting of hexahedral elements. In order to capture the thermal and viscous boundary layers in the fluid, mesh grading is used to obtain a finer mesh at the fluid-solid interface and the leading edge of the plate. The two regions are coupled at the fluid-solid interface with a \texttt{heatTransferInterface} \texttt{regionInterfaceType} which consists of the meshes boundary patches, namely the "bottom" patch from the fluid and the "top" patch from the solid. Table \ref{tab:FOHPcoupledBCs} shows the thermal coupled boundary conditions used in \texttt{multiRegionFoam}. A Dirichlet condition is applied at the fluid side of the fluid-solid interface while a Neumann condition is specified at the solid side. The generic jump and flux boundary conditions (Section \ref{sec:codeCouplingBCs}) are used for partitioned coupling while the region couple patch field \texttt{chtRcTemperature} \cite{noauthor_chtrctemperature_nodate} is used for monolithic coupling. The rest of the boundary conditions are summarized in Table \ref{tab:FOHPBCs}. 
\renewcommand{\tabcolsep}{10pt}
\begin{table}
\centering
\caption{Boundary conditions for the flow over a heated plate}
\begin{tabular}{lccc}
\hline
Boundary        & Thermal & Velocity  \\ \hline
\textbf{Fluid} && \\  \hline
inlet        & 300 K & $(1\;0\;0)^\top\, \unitfrac[]{m}{s} $   \\
bottom     & coupled & $(0\;0\;0)^\top\, \unitfrac[]{m}{s} $    \\
slip-bottom (before the plate)    & zeroGradient & zeroGradient   \\
noSlip-bottom (after the plate)    & zeroGradient & $(0\;0\;0)^\top\, \unitfrac[]{m}{s} $    \\
outlet, top & zeroGradient & zeroGradient \\  \hline
\textbf{Solid} && \\  \hline
top       & coupled & -  \\ 
bottom       & 310 K & -  \\
left, right & zeroGradient  & -  \\
\hline
\end{tabular}
\label{tab:FOHPBCs}
\end{table}
\renewcommand{\tabcolsep}{6pt} 
\begin{table}
\centering
\caption{Coupled thermal boundary conditions for the flow over heated plate simulation}
\begin{tabular}{lccc}
\hline
Region & Boundary & Partitioned   & Monolithic         \\ \hline
Fluid & bottom     & regionCoupledScalarJump & chtRcTemperature       \\
Solid & top        & regionCoupledScalarFlux & chtRcTemperature  \\ \hline
\end{tabular}
\label{tab:FOHPcoupledBCs}
\end{table}
\noindent
The authors of \cite{vynnycky_forced_1998} investigated several factors affecting heat transfer, including the aspect ratio of the plate $\lambda = a/L$, the Reynolds number, Re, the Prandtl number, Pr, and the thermal conductivity ratio, $\operatorname{k}= k_s/k_f$, between the plate and the fluid. In this study, the aspect ratio is fixed at $\lambda = 0.25$, while different combinations of the other parameters are considered as specified in table \ref{tab:FOHPcases}; the thermophysical properties of the fluid and solid are given in table \ref{tab:FOHPParameters}. The simulations are run for $10$s using time step size of $\Delta t = 0.01$. The numerical schemes used are listed in Table \ref{tab:FOHPschemes} according to OpenFOAM equivalent terms \cite{chris_greenshields_openfoam_2022}.
\renewcommand{\tabcolsep}{20pt}
\begin{table}
\centering
\caption{Parameters for the flow over heated plate simulation}
\begin{tabular}{ccc}
\toprule
Re & Pr & k \\ \hline
500 & 0.01 & 1, 5, 20 \\
10000 & 0.01 & 1, 5, 20 \\
500 & 100 & 1, 5, 20 \\
\bottomrule
\end{tabular}
\label{tab:FOHPcases}
\end{table}
\renewcommand{\tabcolsep}{6pt} 
\begin{table}
\centering
\caption{Thermophysical properties of the fluid and solid for the flow over heated plate simulation}
\begin{tabular}{lcccc}
\toprule 
Property & Symbol & Unit & Solid & Fluid \\
\hline
Density & $\rho$ & $\unitfrac[]{kg}{m^3}$ &  $1$ &  $1$ \\
Dynamic viscosity & $\mu$ & $\unitfrac[]{kg}{ms}$ & - & $\rho_f U_{\infty} L / \operatorname{Re}$ \\
Thermal conductivity & $k$ & $\unitfrac[]{W}{m \cdot K}$ & 100 & $k_s / \operatorname{k}$  \\
Specific heat capacity & $c_p$ & $\unitfrac[]{J}{kg \cdot K}$ & 100 & $k_f \operatorname{Pr} / \mu$ \\
\bottomrule
\end{tabular}
\label{tab:FOHPParameters}
\end{table}
\begin{table}
    \centering
    \caption{Numerical schemes for the flow over heated plate simulation}
    \begin{tabular}{lll}
        \hline & Scheme & Setting \\
        \hline
        Time Scheme & ddtScheme & backward \\
        \hline
        Finite Volume Schemes & gradScheme  & leastSquares \\
        & divScheme div(phi,U) & Gauss upwind \\
        & divScheme div(phi,T) & Gauss linearUpwind Gauss linear \\
        & lapacianScheme & Gauss linear corrected \\
        & interpolationScheme & linear \\
        & snGradScheme & corrected \\
        \hline
    \end{tabular}
    \label{tab:FOHPschemes}
\end{table}
\noindent
The results are validated by computing the dimensionless conjugate boundary temperature, $\theta$, defined as  
$$\theta = \frac{T-T_{\infty}}{T_s-T_{\infty}},$$
where $T$ is the temperature along the fluid-solid interface. Figure \ref{fig:FOHPResults} summarizes the results of \texttt{multiRegionFoam} using monolithic and partitioned coupling. The latter was performed with Aitken's relaxation procedure to accelerate the convergence. Both approaches result in good agreement with the numerical and analytical results from Vynnycky et al. \cite{vynnycky_forced_1998}. The deviations from the reference solutions in Figure \ref{fig:FOHPResultsPr100Re500} are due to the fact that the solutions for the case $Pr$ \(\gg\) 1 were derived under the constant-flux approximation assumption, which does not provide an accurate description of the flow as $k$ increases. Figure \ref{fig:FOHPResultsAvgCouplingTime} reports the average time spent on the coupling using partitioned and monolithic approaches. In comparison to partitioned coupling, monolithic coupling exhibits either the same or reduced average coupling time across all cases. However, a notable distinction is observed when Pr is equal to 100. Notably, for monolithic coupling, the time remains almost constant regardless of the simulated parameters.
%

\newpage
\subsection{Shell-and-tube heat exchanger}

\begin{figure}[ht]
\centering
\resizebox{0.8\textwidth}{!}{\input{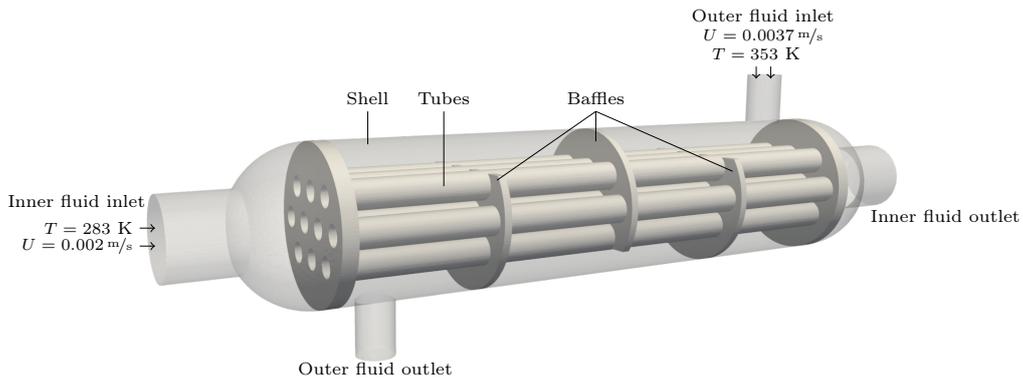}}
\caption{Geometry for the heat exchanger}
\label{fig:HEDomain}
\end{figure}

\begin{figure}[H]
    \centering
    \includegraphics[width=0.8\textwidth]{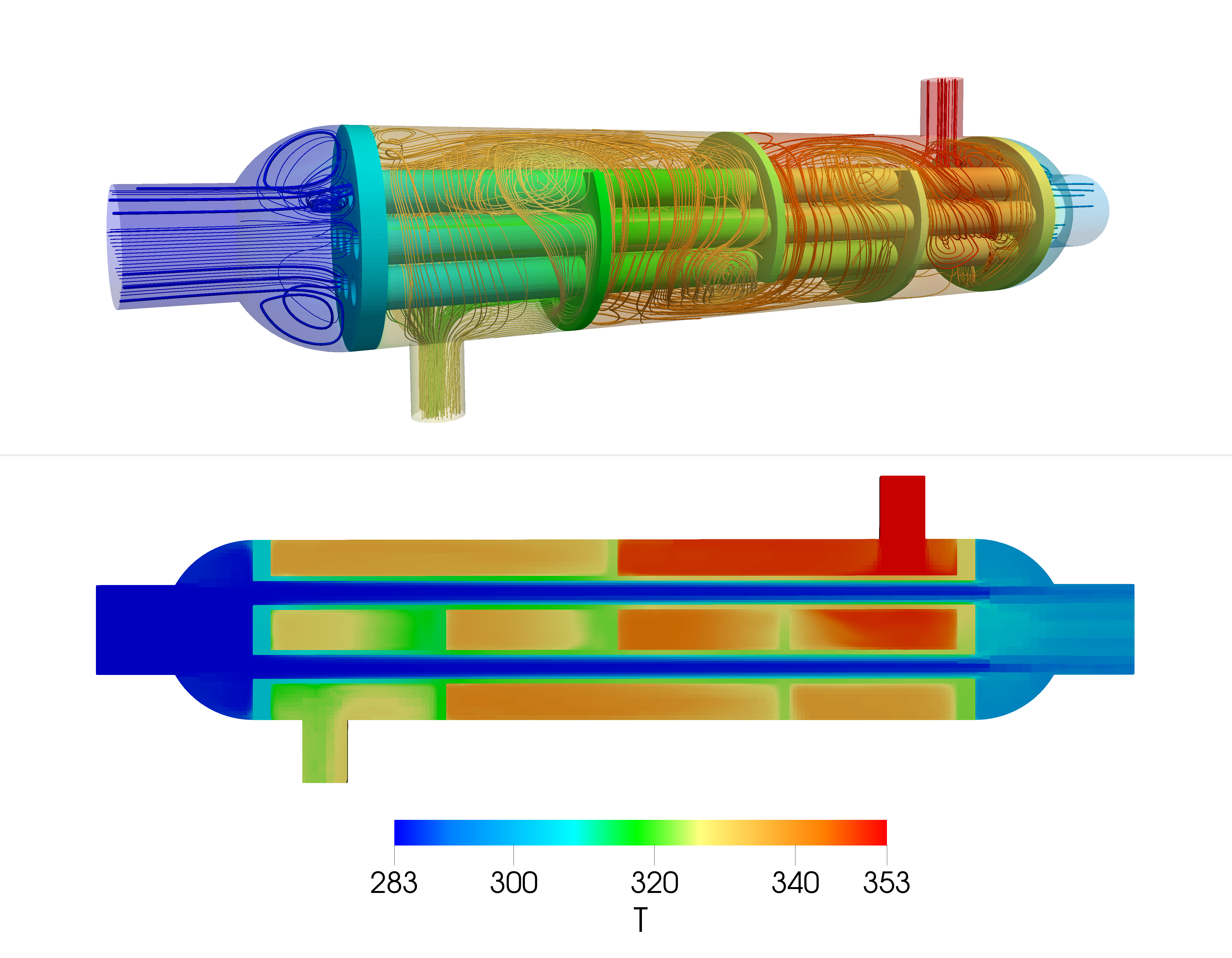}
    \caption{Final temperature distribution for the heat exchanger simulation}
    \label{fig:HEResults}
\end{figure}

The next example demonstrates an industrial application where conjugate heat transfer takes place. Figure \ref{fig:HEDomain} shows a shell-and-tube heat exchanger. This particular design includes a shell, tubes, and baffles. Heat transfer occurs between two fluids; an inner fluid, flowing at lower temperature inside the tubes, and an outer fluid, flowing within the shell, but outside the tubes. The solid walls of the tubes ensure that the two fluids do not mix while the baffles help directing the flow on the shell side.
\begin{figure}[H]
    \centering
    \includegraphics[width=0.8\textwidth]{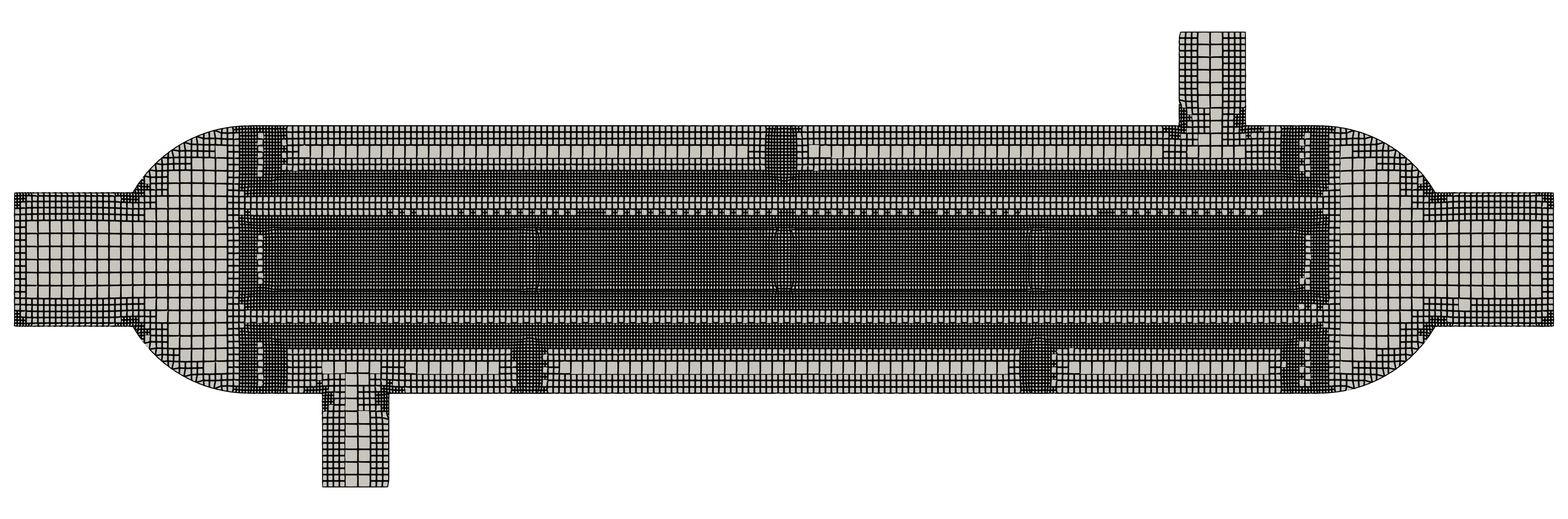}
    \caption{Meshes for the heat exchanger simulation}
    \label{fig:HEmesh}
\end{figure}
\noindent
The case setup is based on a case prepared by SimScale GmbH that is publicly available at \cite{noauthor_heat_nodate}. The  computational mesh is shown in Figure \ref{fig:HEmesh} and the boundary conditions are summarized in Table \ref{tab:HEBCs}. The coupled boundary conditions are as in the previous case in Table \ref{tab:FOHPcoupledBCs}. The material properties of the solid and fluid regions are shown in Table \ref{tab:HEproperties} where the inner and the outer fluids both have the same properties.
\renewcommand{\tabcolsep}{10pt} 
\begin{table}
\centering
\caption{Boundary conditions for the heat exchanger}
\begin{tabular}{lccc}
\toprule
Boundary        & Thermal & Velocity  \\ \hline
\textbf{Inner fluid} && \\ \hline
inlet        & 283 K & $(0.002\;0\;0)^\top\, \unitfrac[]{m}{s} $   \\
inner\_to\_solid     & coupled & $(0\;0\;0)^\top\, \unitfrac[]{m}{s} $   \\
outlet, walls & zeroGradient & zeroGradient \\ \hline
\textbf{Outer fluid} && \\ \hline
inlet        & 353 K & $(0\;0.0037\;0)^\top\, \unitfrac[]{m}{s} $   \\
outer\_to\_solid     & coupled & $(0\;0\;0)^\top\, \unitfrac[]{m}{s} $   \\
outlet, walls & zeroGradient & zeroGradient \\ \hline
\textbf{Solid} && \\ \hline
solid\_to\_inner       & coupled & -  \\ 
solid\_to\_outer       & coupled & -  \\
walls & zeroGradient  & -  \\
\bottomrule
\end{tabular}
\label{tab:HEBCs}
\end{table}
\renewcommand{\tabcolsep}{6pt} 
\begin{table}
\centering
\caption{Thermophysical properties of the fluid and solid for the heat exchanger case}
\begin{tabular}{lcccc}
\toprule 
Property & Symbol & Unit & Fluid & Solid \\
\hline 
Density & $\rho$ & $\unitfrac[]{kg}{m^3}$ & 1027 & 8960 \\
Thermal conductivity & $k$ & $\unitfrac[]{W}{m \cdot K}$ & 0.668  & 401 \\
Dynamic viscosity & $\mu$ & $\unitfrac[]{kg}{ms}$ & $3.645 \mathrm{e}^{-4}$ & - \\
Specific heat capacity & $c_p$ & $\unitfrac[]{J}{kg \cdot K}$ & 4195& 385 \\
\bottomrule
\end{tabular}
\label{tab:HEproperties}
\end{table}
\begin{table}
\centering
\caption{Numerical schemes for the heat exchanger}
\begin{tabular}{lll}
\hline & Scheme & Setting \\
\toprule
Time Scheme & ddtScheme & steadyState \\
\hline
Finite Volume Schemes & gradScheme  & Gauss linear \\
& gradScheme grad(U)   & cellLimited Gauss linear 1 \\
& divScheme div(phi,U) & Gauss upwind \\
& divScheme div(phi,T) & Gauss upwind \\
& lapacianScheme & Gauss linear corrected \\
& interpolationScheme & linear \\
& snGradScheme & corrected \\
\bottomrule
\end{tabular}
\label{tab:HEschemes}
\end{table}
%
\noindent
The simulations are run for $500$s using time step size of $\Delta t = \unit[1]{s}$. The numerical schemes used are listed in Table \ref{tab:HEschemes}. Figure \ref{fig:HEResults} shows the results of the heat exchanger simulation using monolithic coupling. It depicts the distribution of the temperature at t = $\unit[30]{s}$ where no significant change in the temperature is observed afterwards. The computed field values using the partitioned approach are not remarkably different, but it takes longer to reach a steady state (around t = $\unit[500]{s}$).  
%

\subsection{Polymer electrolyte fuel cell}

\begin{figure}[H]
\centering
\includegraphics[clip,width=1\columnwidth, scale=0.2]{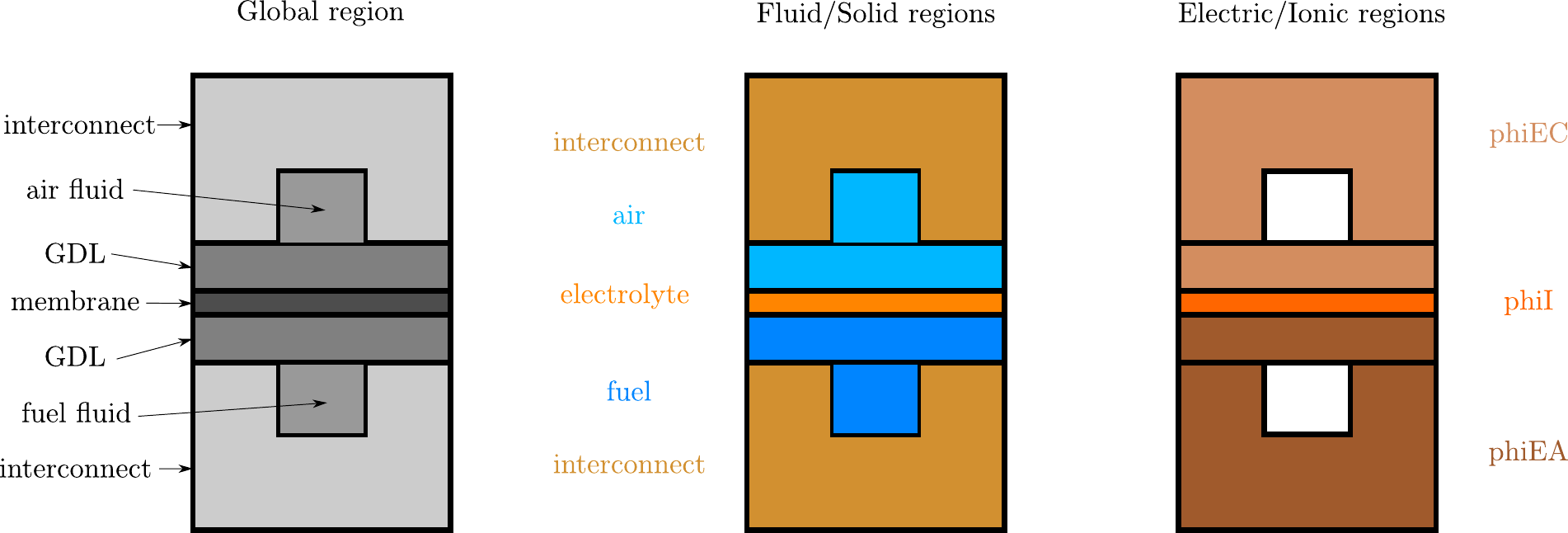}
\caption{General modeling approach of the PEM fuel cell}
\label{fig:fuelCell}
\end{figure}

The possibility of simulating complex multi-physical systems within the \texttt{multiRegionFoam} framework will be demonstrated in the following example, considering a single polymer electrolyte fuel cell (PEMFC) channel. The general structure and the typical physical components of a PEMFC are depicted in the left, gray scaled image of Figure \ref{fig:fuelCell}.   

\begin{figure}[H]
\centering
\includegraphics[clip,width=0.65\columnwidth, scale=0.03]{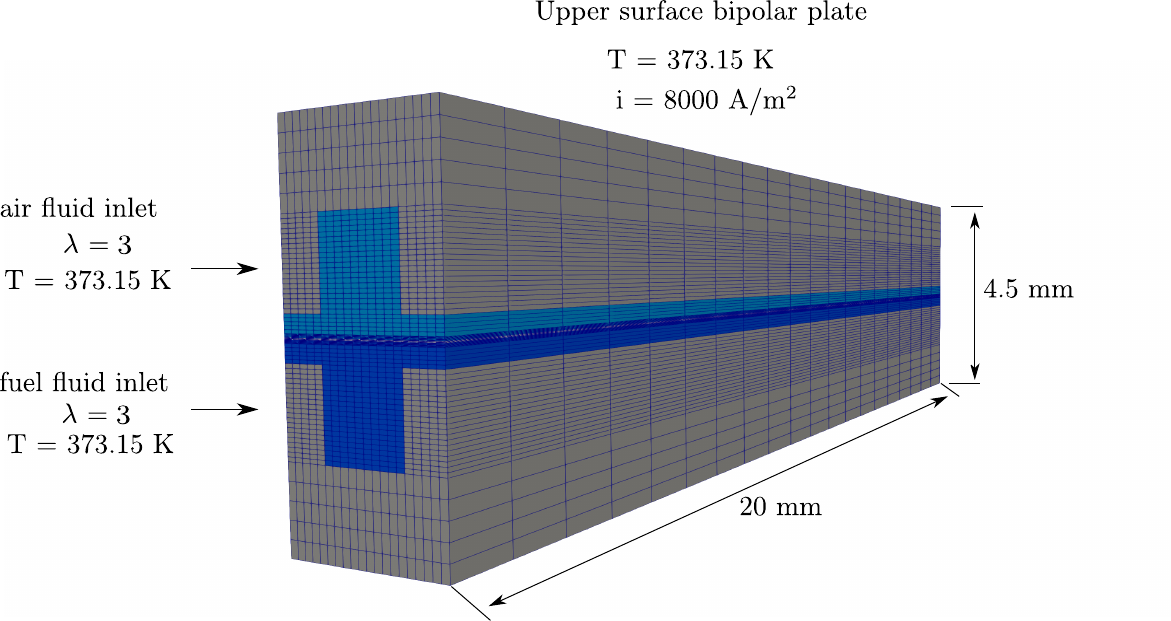}
\caption{Mesh and simulation domain}
\label{fig:fuelCellMesh}
\end{figure}

\noindent
Beginning with the outer components, the cell consists of two end plates, or interconnects, to which the external electrical circuit is connected. The reactants are conveyed through the air and fuel channels and are transported via porous gas diffusion layers (GDLs) to the catalyst layers (CLs), which is represented here as an infinite thin surface. Besides the gas transport to the catalyst layers, the electric conductive GDLs also enable the electrical contact between the forementioned and the end plates (\texttt{interconnect}). The CL is located between the ionic conductive membrane and GDL. On the cathode (CL at the air side) oxygen is reduced and water is created, whereas on the anode side (CL at the fuel side) hydrogen is oxidized and protons are conducted through the ionic conductive membrane to the opposite CL.
In addition, Figure \ref{fig:fuelCell} shows the general structure of the modeling and computational domains. The design of the single-phase PEMFC implemented here in \texttt{multiRegionFoam} is based on developments that can be traced back to \cite{openfuelcell} and to \cite{zhang_pem_2020}, and in this particular case has been extended especially towards the possibility of interface coupling.
Thus, the subdomains defined as different \texttt{regionTypes} are solid, fluid and electric/ionic conductive regions. 
In the fluid (\texttt{air} and \texttt{fuel}) regions, both the modified Navier-Stokes equations, which take into account the porous layers via Darcy's law, and the species mass fraction equations are solved.
The gas mixtures are assumed to be incompressible, ideal gases whose diffusion is characterized by Fick's law, taking also into account the impact of the porous layers. On the cathode side, the inflowing gas mixture is composed of oxygen, nitrogen and water vapor, whereas on the anode side, hydrogen and water vapor flow in. In total, the constraint that the sum of all mass fractions have to equal one must be fulfilled in each cell. Humidification of the gas mixtures is important to ensure the ionic conductivity of the membrane, which is strongly dependent on its water content.
Within the electric/ionic conducting regions (\texttt{phiEC}, \texttt{phiEA} and \texttt{phiI}), the electric potential equations are solved using the partitioned coupling approach. Here, the open cell voltage is described by Nernst equation and the activation overpotential resulting from the reaction is expressed by Butler-Volmer equation. Additional contact resistances or limiting current densities are not considered in this particular case.  
The cell operates in galvanostatic mode, i.e.,\ a constant mean current density is applied to it. Therefore, the potential field specified at the upper surface of \texttt{phiEC} (interconnectUp) is adjusted every iteration step until the specified current density is reached.
In Table \ref{tab:FuelBCs} this is indicated by the entry \texttt{adjustedPotential} for the boundary interconnectUp.   
The calculation of the temperature within each component of the PEM fuel cell takes place on the global region by mapping the required quantities such as densities and specific heat capacities etc. onto this mesh. A further adaptation of the source code, whereby the energy equation on the individual regions are solved using the coupled boundary conditions as shown in the previous sections, is also be possible with the framework here. With such an approach and the assumption of a single-phase flow, the global region would be redundant.
Figure \ref{fig:fuelCellMesh} shows the dimensions of the single channel and selected boundary conditions taken from Table \ref{tab:FuelBCs}.  

\begin{figure}[H]
\centering
\subfloat[Velocity distribution
\label{Fig:velHTPEM}]{\includegraphics[width=0.63\textwidth,trim={0 15cm 0 15cm},clip]{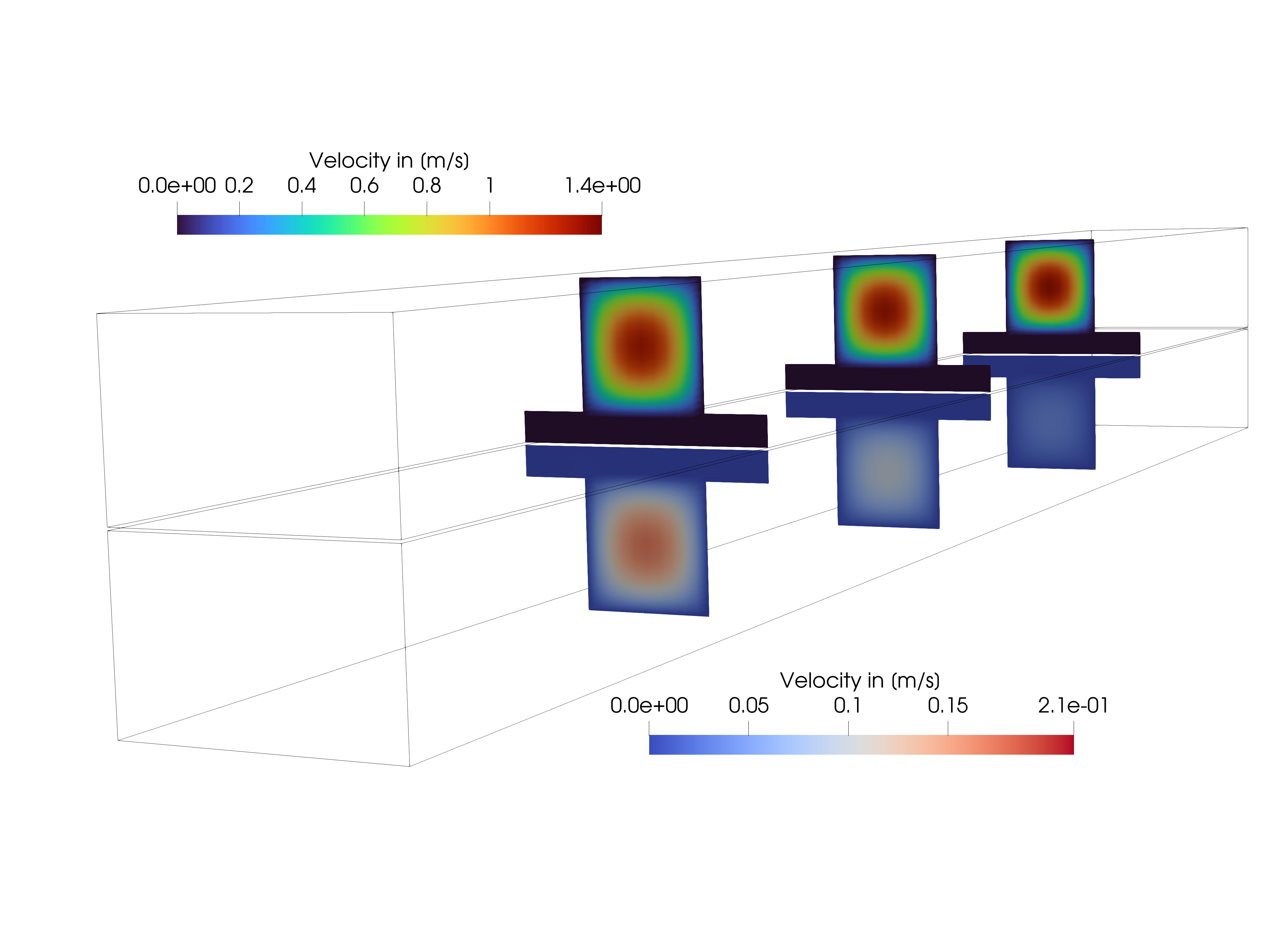}}\hfill
\subfloat[Species mass fractions \label{Fig:speciesHTPEM}]{\includegraphics[width=0.63\textwidth,trim={0 15cm 0 15cm},clip]{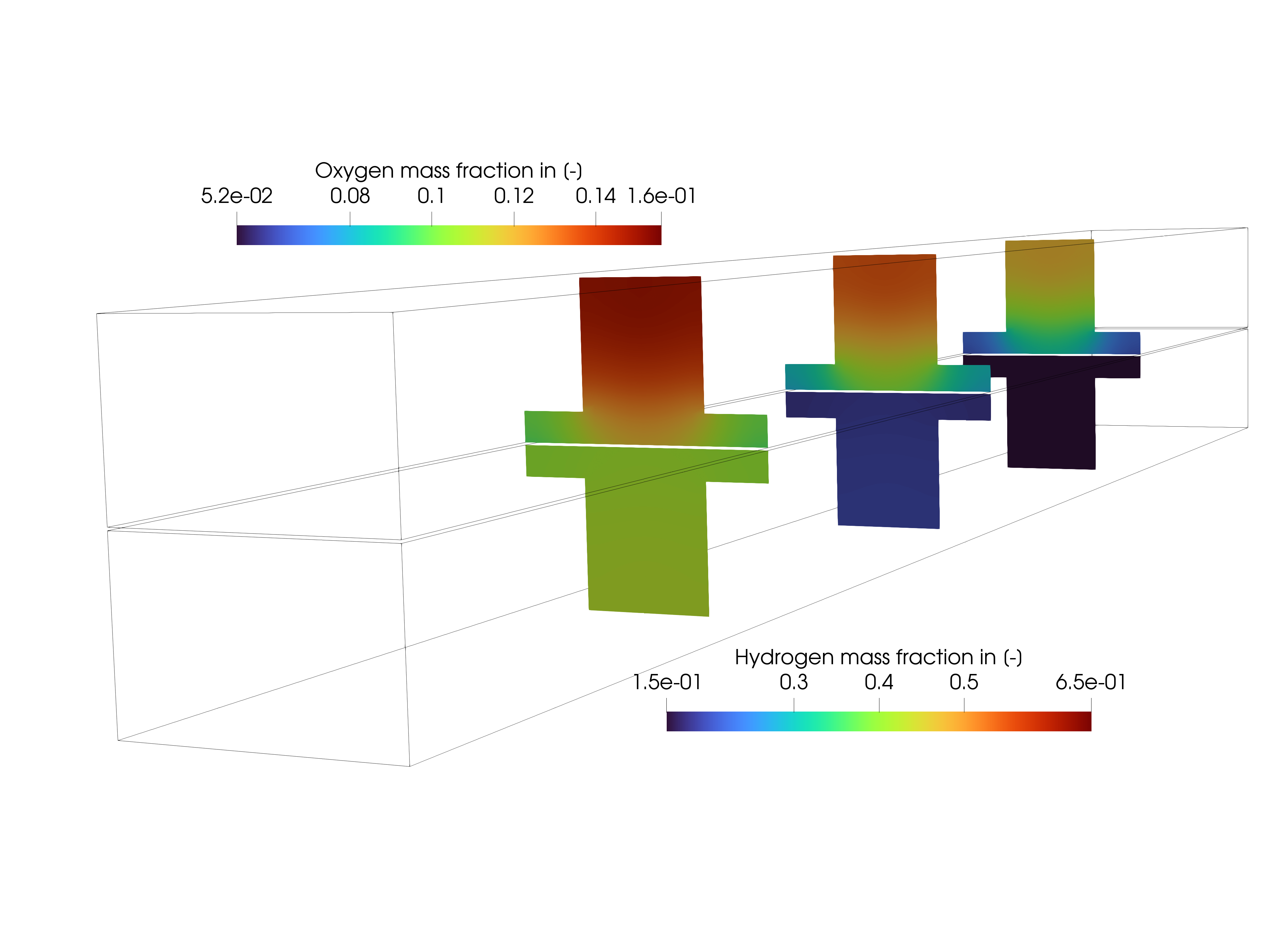}}
\caption{Velocity and species mass fraction distribution along the channels for $i = \unitfrac[8000]{A}{m^2}$}
\label{Fig:resHTPEM}
\end{figure}

\begin{table}
\centering
\caption{Boundary conditions for the PEM fuel cell}
\begin{tabular}{lcccc}
\toprule
Boundary        & Thermal & Velocity & Species mass fraction & Potential  \\ \hline
\textbf{air} && \\ \hline
airInlet        & - & $(0.655 \, 0 \, 0)^\top \, \mathrm{m/s}$ & $\mathrm{Y_{O2}}$ = 0.156, $\mathrm{Y_{H2O}}$ = 0.35 & -    \\
airOutlet     & - & zeroGradient & zeroGradient & -   \\
air\_to\_interconnect & - & noSlip & zeroGradient & - \\
air\_to\_electrolyte & - & noSlip & zeroGradient & - \\
airSides & - & noSlip & zeroGradient & - \\ \hline
\textbf{fuel} && \\ \hline
fuelInlet        & - & $(0.161 \, 0 \, 0)^\top \, \mathrm{m/s}$ & $\mathrm{Y_{H2}}$ = 0.65, $\mathrm{Y_{H2O}}$ = 0.35 & -    \\
fuelOutlet     & - & zeroGradient & zeroGradient & -   \\
fuel\_to\_interconnect & - & noSlip & zeroGradient & - \\
fuel\_to\_electrolyte & - & noSlip & zeroGradient & - \\
fuelSides & - & noSlip & zeroGradient & - \\ \hline
\textbf{phiI} && \\ \hline
electrolyteSides        & - & - & - & zeroGradient   \\
electrolyte\_to\_air     & -  & - & - &  coupled flux   \\
electrolyte\_to\_fuel & - & -& - & coupled jump \\ \hline
\textbf{phiEA} && \\ \hline
anodeSides        & - & - & - & zeroGradient   \\
interconnectDown    & - & - & - & 0 V   \\
phiEA\_to\_electrolyte  & - & - & - & coupled flux   \\ \hline
\textbf{phiEC} && \\ \hline
cathodeSides        & - & - & - & zeroGradient   \\
interconnectUp    & - & - & - & adjustedPotential   \\
phiEC\_to\_electrolyte  & - & - & - & coupled jump   \\ \hline
\textbf{Global region} && \\ \hline
airInlet        & 363.15 K & - & - & -   \\
airOutlet     & zeroGradient & - & - & -   \\
fuelInlet        & 363.15 K & - & - & -   \\
fuelOutlet     & zeroGradient & - & - & -   \\
interconnectSides & zeroGradient & - & - & -   \\
interconnectUp        & 363.15 K & - & - & -   \\
interconnectDown     & 363.15 K & - & - & -   \\ 
\bottomrule
\end{tabular}
\label{tab:FuelBCs}
\end{table}
\noindent
The cell under consideration consists of a single channel with a length of 20 \,mm, where the fluid channels have an area of 1 $\times$ 1 $\mathrm{mm}$. The simulation itself is steady state and the velocity used for the boundary condition at the fluid inlets is given by a constant stoichiometry of $\lambda = 3$ calculated via Faraday's law.
The temperature of the incoming gases is set to $T = 90 \, \mathrm{^{\circ}C}$ and a mean current density of $i = 8000 \, \mathrm{A/m^{2}}$ is applied. Figure \ref{Fig:resHTPEM} shows the corresponding velocity and species mass fraction contours at different positions along the channel inside the anodic and cathodic fluid regions.

\begin{figure}[H]
\centering
\subfloat[Potential distribution in the middle of the cell from the top of the cathode to the buttom of the anode GDL for 
$i$ = 6000\! $\mathrm{A/m^2}$
\label{Fig:potJump}]{\includegraphics[width=0.5\textwidth]{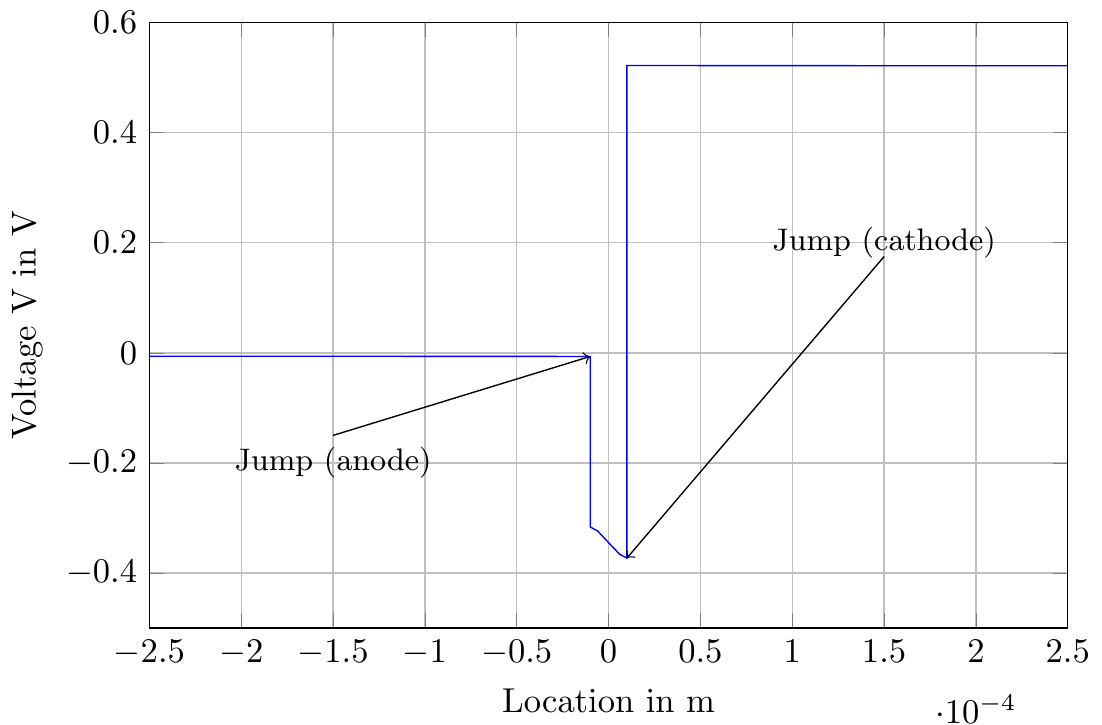}}\hfill
\subfloat[Polarization curve \label{Fig:polar}]{\includegraphics[width=0.479\textwidth]{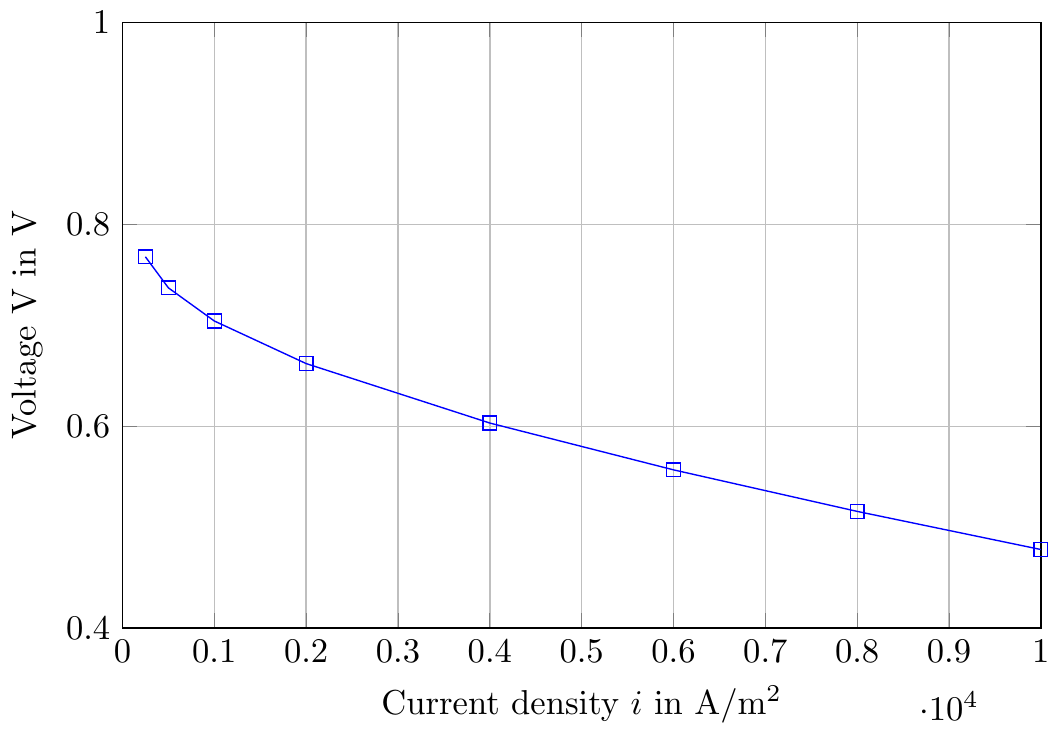}}
\caption{Potential distribution along the middle of the cell (a) and the polarization curve (b)}
\label{Fig:scaling}
\end{figure}

\noindent
At the air side the velocity increases due to the combination of the reaction and decreased density of the gas mixture, dragged water and the development of the velocity profile, since a constant velocity is specified at the inlet. 
For the fuel side, the velocity is decreasing, mainly due to the dragged water and the consumption of hydrogen. 
This corresponds to the mass fraction contours, where the oxygen and hydrogen mass fraction decreases in flow direction.
The main focus within this example is in the surface coupling of the potential. 
The reactions take place on the infinitesimally thin CL and at this interface the coupled jump/flux boundary conditions for the potential are prescribed (see Table \ref{tab:FuelBCs}). 
Figure \ref{Fig:potJump} shows the potential variation in the middle of the cell going from the upper surface of the GDL at the cathode to the lower surface of the GDL at the anode.
Due to the high electric conductivity of the porous media, the potential drop there is comparatively low. At the CLs however, the resulting overpotentials arising from the reactions induce a jump in the potential. 
Another noticeable potential drop can be found in the membrane because of the low ionic conductivity of the membrane.
In addition to this, Figure \ref{Fig:polar} shows the polarization curve as a result of conducting the simulation for different current densities.
The polarization curve of a PEM fuel cell exhibits a characteristic pattern, featuring a non-linear region at low current densities, which is primarily influenced by the activation overpotential, and a subsequent linear ohmic region, predominantly governed by the internal resistances within the cell.    

%

\subsection{Air bubble rising in water}
\label{sec:RB}
\begin{figure}[H]
    \centering
    \includegraphics[width=\textwidth]{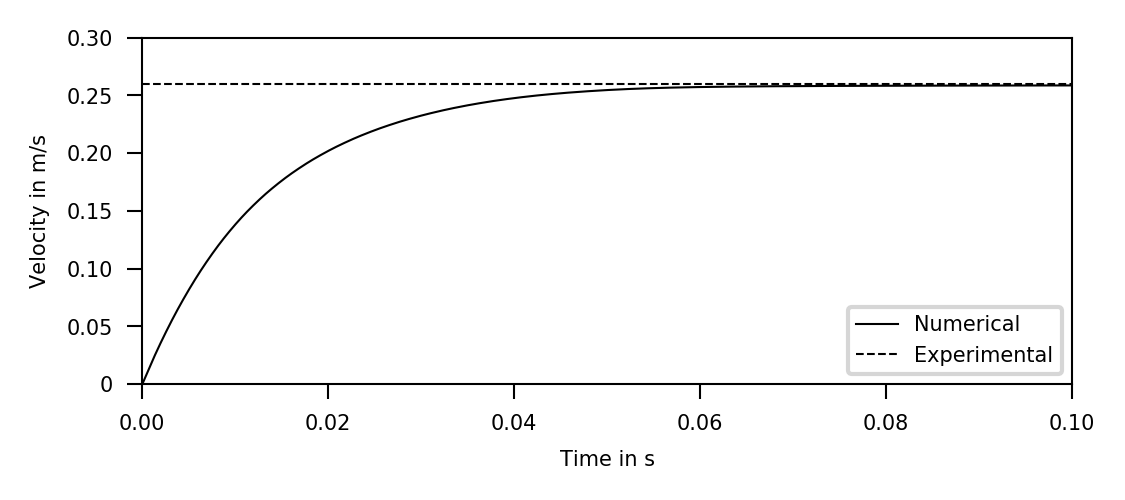}
    \caption{The rise velocity over time for bubble with radius $r_b=$ $\unit[0.5]{mm}$ compared with the experiment results}
    \label{fig:RB05riseVelocity}
\end{figure}

\begin{figure}[H]
    \centering
    \includegraphics[width=\textwidth]{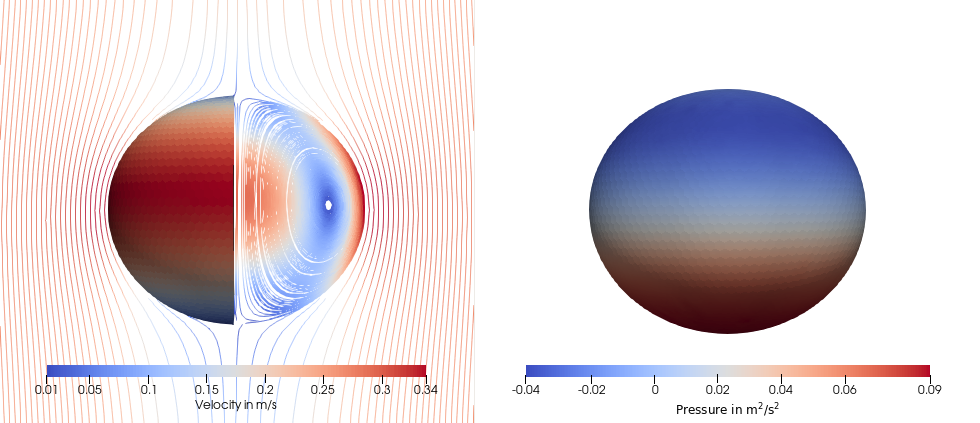}
    \caption{Velocity and pressure fields visualisation at time $\unit[0.1]{s}$}
    \label{fig:RB05BubbleView}
\end{figure}

\begin{figure}[H]
    \centering
    \includegraphics[width=\textwidth]{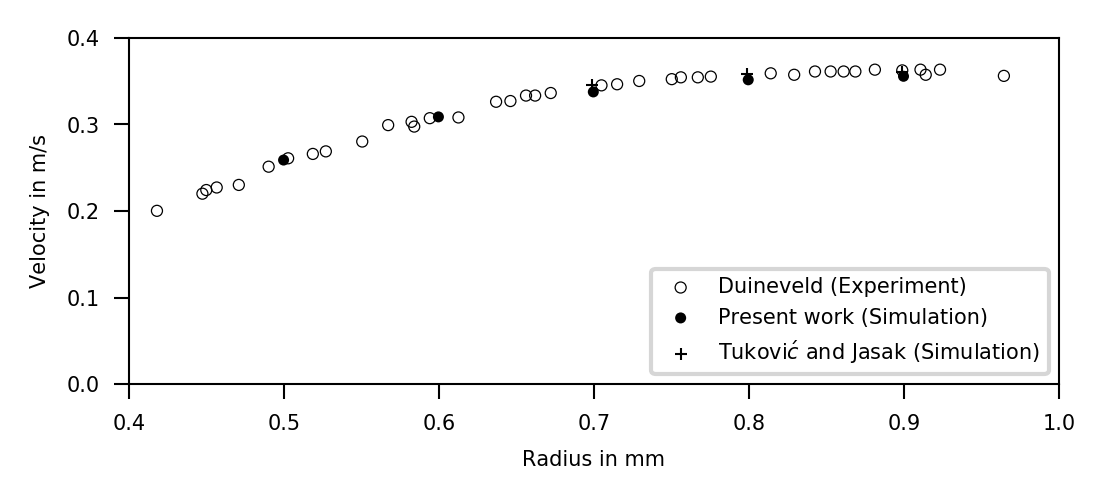}
    \caption{Comparison of the terminal rise velocity for varying equivalent bubble radii} 
    \label{fig:RBRadiusVelocity}
\end{figure}

\begin{figure}[H]
    \centering
    \includegraphics[width=0.8\textwidth]{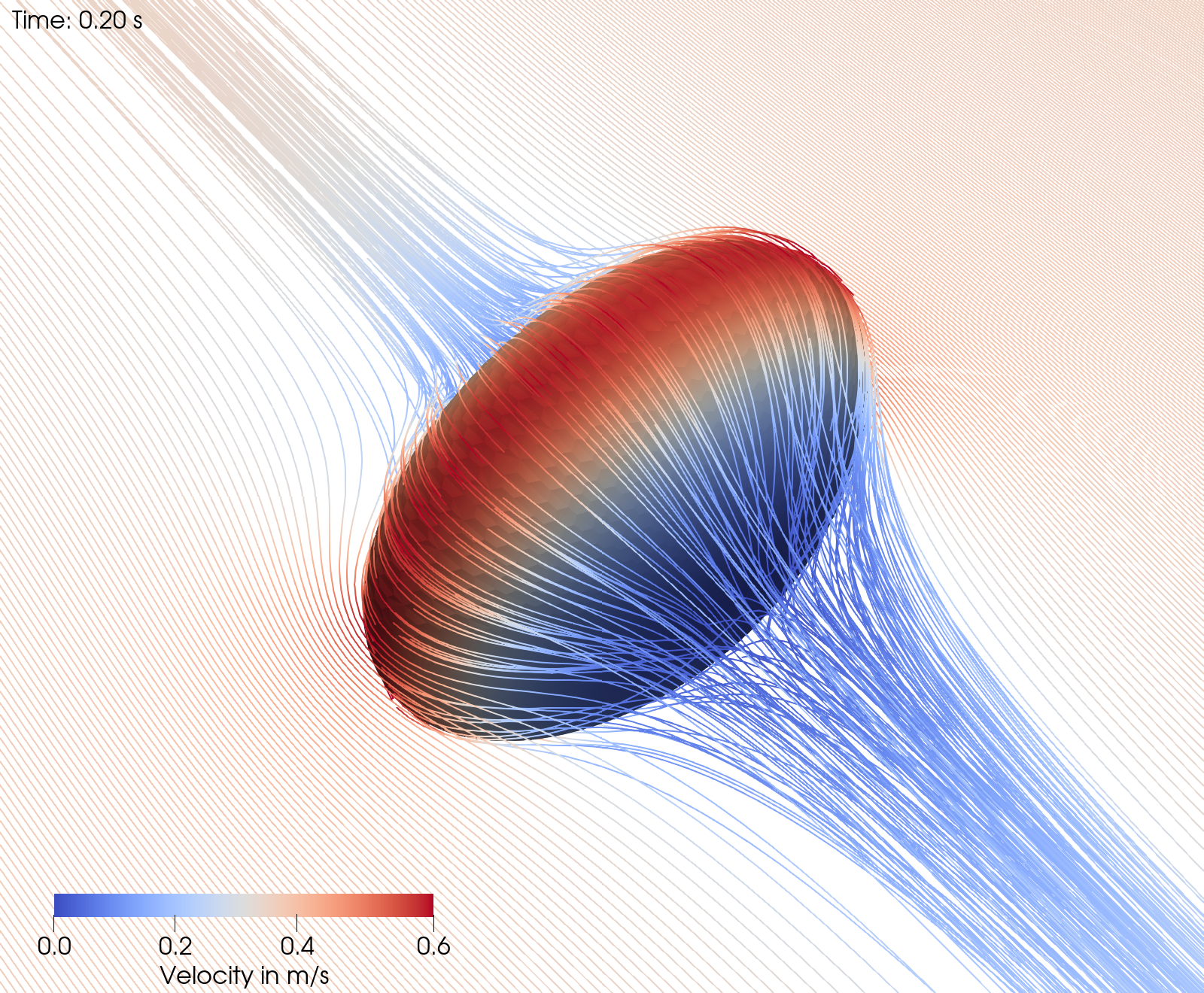}
    \caption{Velocity field visualisation for the rising bubble with $r_b = 0.9$} 
    \label{fig:RB09Velocity}
\end{figure}

The last two cases demonstrate the usage of \texttt{multiRegionFoam} to implement a moving mesh ALE interface tracking method, originally developed by Hirt et al. \cite{Hirt_ALE} and originally implemented into OpenFOAM by Tukovi\'c and Jasak \cite{tukovic_moving_2012}. A single air bubble rising in still pure water is considered, following the setup of Duineveld \cite{Duineveld1995} and using his experimental data for validation. The bubble assumes an initial spherical shape with radius $r_b$ and  accelerates from zero velocity at release to its terminal rise velocity. As shown in Figure \ref{fig:RBDomain}, the computational domain has two regions comprising the bubble and the outer medium which is represented by a sphere of radius $20r_b$. The bubble mesh is bounded by the "interfaceShadow" patch which coincides with the "interface" patch from the liquid side forming the bubble-liquid interface where the coupled boundary conditions are imposed according to Table \ref{tab:RBBCs}. The meshes consist of polyhedral cells for the bubble and prismatic cells with a polyhedral base for the water, as shown in Figure \ref{fig:RBmesh}.
\begin{figure}[htp]
\centering
    \begin{tikzpicture}

\draw[color=black](0,0) circle (3.2); 

\draw[color=black](0,0) circle (1); 

\draw[dashed,color=black](0,0) circle (0.95);

\node at (0,0.2) {FluidB};
\node at (0,2) {FluidA};

\draw[-] (0,0) -- (1,0) node[right] {$r_{_b}$};

\draw[-] (0.5,-0.89) -- (1.5,-1.4) node[below] {interface};

\draw[-] (-0.38,-0.85) -- (-1,-1.4) node[below] {interfaceShadow};

\draw[-] (3.05,-1)  -- (3.5,-1.4) node[below] {space};

\draw[->] (-3,-2.4) -- (-3,-2.8) node[below] {$\mathbf{g}$};

\end{tikzpicture}
\caption{Sketch of the computational domain for the rising bubble simulation}
\label{fig:RBDomain}
\end{figure}
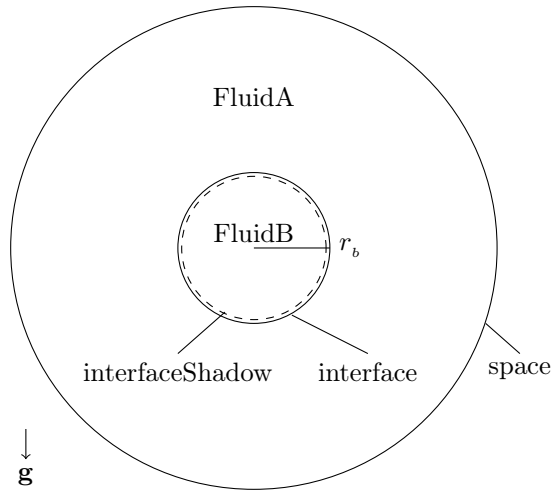
\begin{figure}[H]
\centering
\resizebox{1\textwidth}{!}{\input{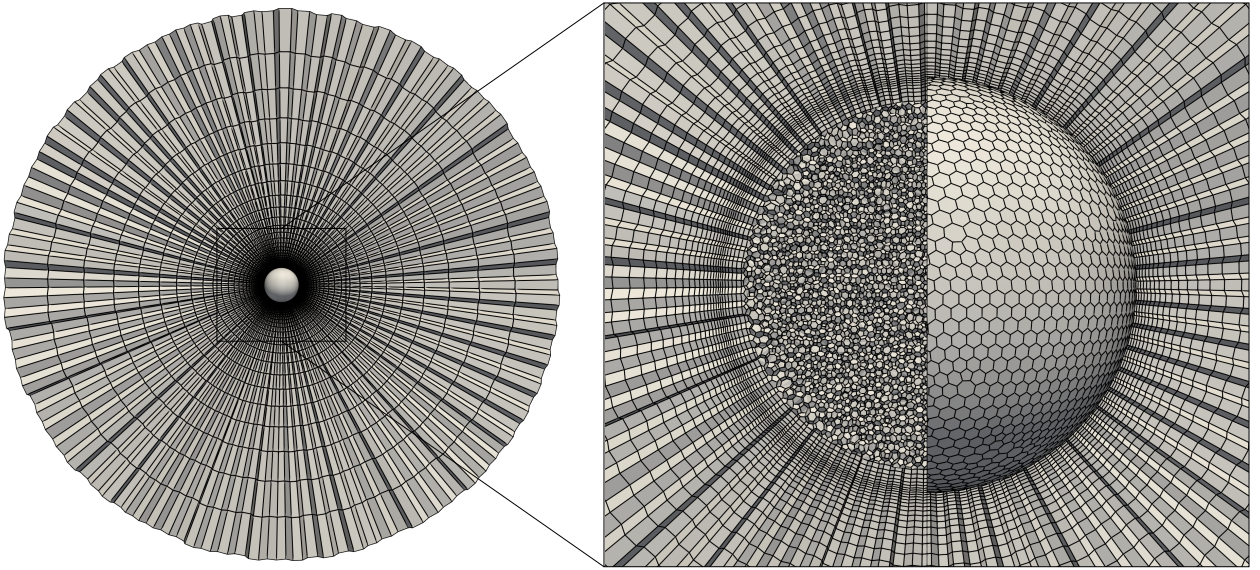}}
\caption{Mesh for the rising bubble simulation}
\label{fig:RBmesh}
\end{figure}
\begin{table}[h]
    \centering
    \caption{Boundary conditions for the rising bubble simulation}
    \begin{tabular}{lccc}
    \toprule
    Boundary        & Pressure & Velocity  \\ \hline
    \textbf{FluidA} && \\  \hline
    interface   & genericInterfaceCoupledPressureValue & genericInterfaceCoupledVelocityFlux   \\
    space & zeroGradient & inletOutlet \\  \hline
    \textbf{FluidB} && \\  
    \hline
    interfaceShadow       & genericInterfaceCoupledPressureFlux & genericInterfaceCoupledVelocityValue  \\ 
    \bottomrule
    \end{tabular}
    \label{tab:RBBCs}
\end{table}
\noindent The study considered bubbles with equivalent initial radii of $r_b =  0.5, \, 0.6, \, 0.7, \, 0.8, \, \text{and } \unit[0.9]{mm}$. The physical properties of the bubble and the surrounding liquid are reported in Table \ref{tab:RBfluidProperties}. The simulations run with time step size of $\Delta t = \unit[1e^{-5}]{s}$ until the terminal rise velocities are observed. The numerical schemes used are listed in Table \ref{tab:FOHPschemes}. 
\renewcommand{\tabcolsep}{12pt}
\begin{table}[htp] 
\centering
\caption{Physical properties for the rising bubble simulation}
\begin{tabular}{lll}
\toprule 
Property &  FluidB (Air) & FluidA (Water) \\ \hline
Density & $\unitfrac[1.205]{kg}{m^3}$ & \unitfrac[$998.3]{kg}{m^3}$ \\
Dynamic viscosity & $\unitfrac[1.82e^{-5}]{kg}{ms}$ & $\unitfrac[1e^{-3}]{kg}{ms}$ \\ 
Surface tension coefficient & \multicolumn{2}{c}{$\unitfrac[0.0727]{N}{m}$}\\
\bottomrule
\end{tabular}
\label{tab:RBfluidProperties}
\end{table}
\renewcommand{\tabcolsep}{6pt} 
\begin{table}[htp]
    \centering
    \caption{Numerical schemes for the rising bubble simulation}
    \begin{tabular}{lll}
        \toprule 
        & Scheme & Setting \\
        \hline
        Time Scheme & ddtScheme & backward \\
        \hline
        Finite Volume Schemes & gradScheme  & Gauss linear \\
        & divScheme div(phi,U) & Gauss GammaVDC 0.5 \\
        & divScheme div(phi,T) & Gauss linearUpwind Gauss linear \\
        & lapacianScheme & Gauss linear corrected \\
        & interpolationScheme & linear \\
        & snGradScheme & corrected \\
        \hline
        Finite Area Schemes & gradScheme  & Gauss linear \\
        & divScheme & Gauss linear \\
        & interpolationScheme & linear \\ 
    \bottomrule
    \end{tabular}
    \label{tab:RBschemes}
\end{table}
The results are compared with the experimental data obtained by Duineveld \cite{Duineveld1995}, as well as the simulation results from Tukovi\'c and Jasak \cite{tukovic_moving_2012}. 
Figure \ref{fig:RB05riseVelocity} depicts the rise velocity of the bubble with radius $r_b=\unit[0.5]{mm}$ over time until the expected terminal rise velocity value is attained. Figure \ref{fig:RB05BubbleView} shows the final shape of the bubble at time $\unit[0.1]{s}$ color-coded by the magnitude of velocity and pressure fields along with the velocity streamlines of the flow surrounding the bubble and inside it. Figure \ref{fig:RBRadiusVelocity} displays the terminal rise velocity for larger bubbles. The high level of agreement between the simulation results and the available data from the literature indicates that the present work has successfully addressed any potential robustness issues related to significant mesh deformation. Figure \ref{fig:RB09Velocity} depicts the terminal state of the rising bubble with $r_b=\unit[0.9]{mm}$.
%
\subsection{Air bubble oscillating in water}
\label{sec:OB}
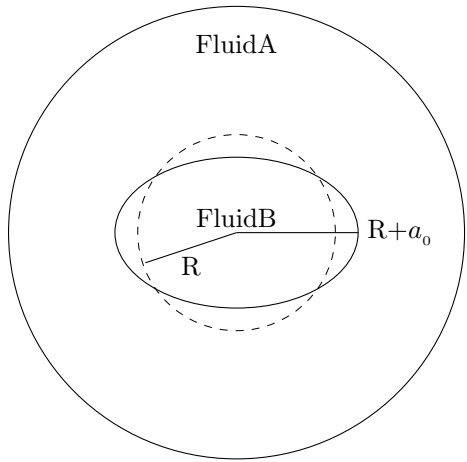
\begin{figure}[h]
\centering
    \begin{tikzpicture}

\draw[color=black](0,0) circle (3); 

\draw (0,0) ellipse (1.6cm and 1.cm);

\draw[dashed,line width=.1mm,color=black](0,0) circle (1.3);

\node at (0,0.2) {FluidB};
\node at (0,2.5) {FluidA};

\draw[-] (0,0) -- (-1.2,-0.4) node[midway,below] {R};

\draw[-] (0,0) -- (1.6,0) node[right] {R$+a_{_0}$};

\end{tikzpicture}
\caption{Sketch of the computational domain for the oscillating bubble simulation }
\label{fig:OBDomain}
\end{figure}

\begin{figure}[H]
    \centering
    \includegraphics[width=0.9\textwidth]{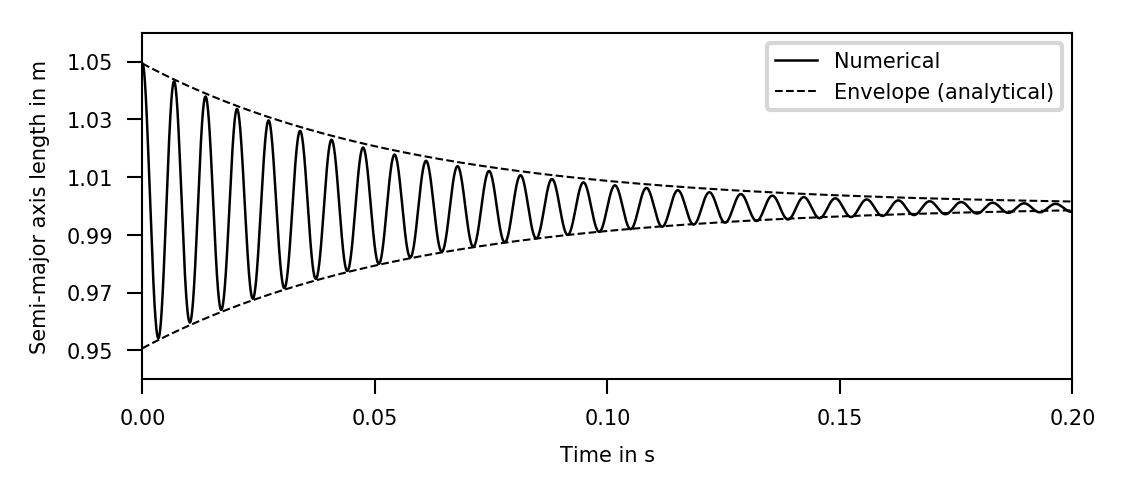}
    \caption{Time evolution of the oscillating bubble compared to the analytical decay profile}
    \label{fig:OBResults}
\end{figure}

Another typical benchmark problem for interface tracking codes involves an air bubble oscillating in a liquid due to interfacial tension in the absence of gravitational forces. The bubble is initially slightly deformed from the shape of a sphere with radius $R$. This results in an initial prolate shape with a semi-major axis $R+a_{_0}$ as shown in Figure \ref{fig:OBDomain}. The boundary conditions, the mesh, and the numerical schemes are the same as in the rising bubble case (Section \ref{sec:RB}). The simulation parameters are summarized in Table \ref{tab:OBparameters} as suggested in \cite{LALANNE_oscillating_2012}. The shape oscillation of the bubble is realised by tracing the temporal evolution of the semi-major axis as illustrated in Figure \ref{fig:OBResults}. The results exhibit high agreement with the analytical decay profile which is given for small linear oscillations by $a_{_0}e^{-t/\tau}$, where $\tau^{-1}$ is the decay factor defined in \cite{prosperetti1980}. 
\begin{table}[h]
\caption{Simulation parameters for the oscillating bubble simulation}
\label{tab:tv1}
\centering
\begin{tabular}{ccccccc}
\toprule
$\rho_{_A}$  & $\rho_{_B}$ & $\mu_{_A} $  & $\mu_{_B} $ & $\sigma $ & $R$ & $a_{_0}$ \\
\midrule
$\unitfrac[1000]{kg}{m^3}$ & $ \unitfrac[1.226]{kg}{m^3}$ & $  \unitfrac[1.13e^{-3}]{kg}{ms}$ & $  \unitfrac[1.78e^{-5}]{kg}{ms} $ & $ \unitfrac[0.0727]{N}{m}$ & $\unit[1]{m}$ & $ \unit[0.05]{m}$\\
\bottomrule
\label{tab:OBparameters}
\end{tabular}
\end{table}


\section{Summary and Outlook}
The present contribution sets out \texttt{multiRegionFoam}, a unified framework for multiphysics problems of  multi-region coupling type within OpenFOAM (FOAM-extend). 
\texttt{multiRegionFoam} offers a flexible design that allows for the assembly of multiphysics problems region-by-region and the specification of coupling conditions interface-by-interface. 
For this, the framework enables to formulate region-specific physics in form of sets of partial differential equations in a modular fashion and  incorporates mathematical jump/transmission conditions in their most general form -- accommodating tensors of any rank. This advancement allows for a unified treatment of coupled transport processes across regions. Moreover, users have the freedom to choose between monolithic and partitioned coupling for each coupled transport equation separately. To address fluid flow problems, the code implements various pressure-velocity algorithms, including SIMPLE, PISO, and PIMPLE, incorporating loops of predictor and corrector steps across regions. 
The framework's maturity and versatility is demonstrated through its successful deployment in various multi-region coupling cases, including multiphase flows, conjugate heat transfer, and fuel cells.
\noindent
The authors anticipate that the creation and release of \texttt{multiRegionFoam} will empower numerous domain experts and developers to derive significant advantages from this framework. Specifically, they aspire for this contribution to facilitate thorough investigations in diverse domains of multiphysics problems and across various application areas. This, in turn, is expected to foster synergistic collaborations among previously separate disciplines, enabling mutual benefits to be realized.

\section*{Acknowledgment}
The authors H.A.\, M.E.F.\ and H.M. are grateful for the funding by the Hessian Ministry of Higher Education, Research, Science and the Arts, and the National High Performance Computing Center for Computational Engineering Science (NHR4CES). S.H.\ is funded by the AI Data Analytics and Scalable Simulations (AIDAS) project, the financial support of which is highly appreciated. Computations for this work were partly conducted on the Lichtenberg II high performance computer of the
Technical University of Darmstadt.

\bibliography{multiRegionFoam}

\end{document}